\documentclass[journal ,12]{IEEEtran}
\usepackage{colortbl}
\usepackage[table]{xcolor}

\usepackage{tikz}
\usepackage{url}
\usepackage{float} 
\usepackage{amsmath,amsfonts,amssymb,mathrsfs,amsthm,amsbsy,graphicx,color,bm,balance,mathtools,xfrac,nccmath,siunitx,cite,soul}
\usepackage{multirow}

\newtheorem{rem}{Remark}

\usepackage[font=scriptsize,labelfont=bf]{caption}
\DeclareMathOperator{\Tr}{\rm{Tr}}

\usepackage{pgfplots}
\pgfplotsset{compat=newest}
\usetikzlibrary{plotmarks}
\usetikzlibrary{arrows.meta}
\usepgfplotslibrary{patchplots}
\usepackage{grffile}
\usetikzlibrary{plotmarks}

\usepackage{pifont}
\newcommand{\cmark}{\ding{51}}%
\newcommand{\xmark}{\ding{55}}%

\usepackage{algpseudocode}
\usepackage{algorithm}
\newcommand{\bc}{\text{BackCom}\xspace}

\newcommand{\abc}{\text{AmBC}\xspace} 
 \usepackage[utf8]{inputenc}
\usepackage{booktabs, caption, makecell,xspace}

\usepackage{threeparttable}

\newcommand{\Exp}[1]{\mathrm{e}^{#1}}

\definecolor{ambe}{rgb}{1.0, 0.49, 0.0}

 \theoremstyle{definition}
 \makeatletter
\newcommand{\printfnsymbol}[1]{%
  \textsuperscript{\@fnsymbol{#1}}%
}
\makeatother
\begin{document}
\bstctlcite{IEEEexample:BSTcontrol}

%
\title{NOMA-Assisted Symbiotic Backscatter: Novel Beamforming Designs Under Imperfect SIC }
\author{Fatemeh Rezaei\textsuperscript{\textasteriskcentered}\thanks{\printfnsymbol{1}F. Rezaei and D.   Galappaththige contributed equally to this work.}, \IEEEmembership{Member, IEEE}, Diluka   Galappaththige\printfnsymbol{1}, \IEEEmembership{Member, IEEE},   Chintha Tellambura, \IEEEmembership{Fellow, IEEE,}   Sanjeewa Herath, \IEEEmembership{Member, IEEE}
\thanks{F. Rezaei, D.  Galappaththige, and C. Tellambura are with the Department of Electrical and Computer Engineering, University of Alberta, Edmonton, AB, T6G 1H9, Canada (e-mail: {rezaeidi, diluka.lg, ct4}@ualberta.ca).  \\
\indent S. Herath is with Huawei Canada, 303 Terry Fox Drive, Suite 400, Ottawa, Ontario K2K 3J1 (e-mail: sanjeewa.herath@huawei.com).}  }


\maketitle

\begin{abstract}
Optimal beamforming designs under imperfect successive interference cancellation (SIC) decoding for a symbiotic network of non-orthogonal multiple access (NOMA) primary users and a secondary ambient tag have been lacking. We address that issue here.  The primary base station (BS) serves  NOMA users and a passive tag simultaneously in this network.    We develop two transmit beamforming designs to meet the user and tag requirements while mitigating the effect of imperfect SIC. Specifically, we design optimal BS  transmit beamforming and power allocation to either maximize the weighted sum rate of  NOMA users and the tag or minimize the BS transmit power under the minimum rate requirements while satisfying the tag’s minimum energy requirement. Because both these problems are non-convex, we propose algorithms using alternative optimization, fractional programming, and semi-definite relaxation techniques. We also analyze their computational complexity. Finally, we present extensive numerical results to validate the proposed schemes and to show significant performance gains while keeping the tag design intact. For example, the proposed digital beamforming increases the harvested power and data rate by \qty{2.16e3}{\percent} and  \qty{314.5}{\percent} compared to random beamforming. 
\end{abstract}


\IEEEpeerreviewmaketitle
\section{Introduction}

\subsection{What Are  Symbiotic Backscatter Networks?} 
\begin{figure}[!t]\centering \vspace{-0mm}
    \def\svgwidth{200pt} 
    \fontsize{8}{8}\selectfont 
    \graphicspath{{Figures/}}
    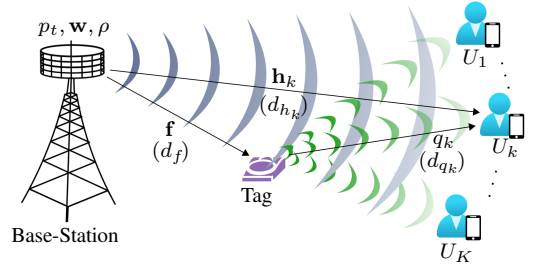 
    \caption{Multiple users supported with single backscatter tag.  }\label{fig:SystemFig}
\end{figure}

Symbiotic radio (SR) is emerging as a means to enhance spectral efficiency (SE) and energy efficiency (EE) 
 of wireless networks, necessitated by the exponential growth of wireless data traffic, devices, and applications.   In biology,  symbiosis is the interaction between two dissimilar organisms living in close physical association. SR, as a strategy to promote the win-win growth of SE and EE, enables primary and secondary networks to cooperate in bandwidth, power, and other resource sharing \cite{Liang2022,Guo2019,Long2020}. For example,   smart home sensors (tags) in an   Internet-of-Things (IoT) network can live together with a cellular network   (Fig.~\ref{fig:SystemFig}).  Both positive (beneficial) and negative (harmful) associations are therefore included. 

\begin{figure}[!tbp]
   \centering
    \def\svgwidth{260pt} 
    \fontsize{8}{8}\selectfont 
    \graphicspath{{Figures/}} 
    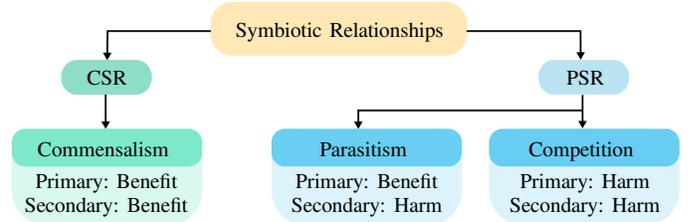 
    \caption{Symbiotic relationships between primary \& secondary networks.}
    \label{fig:symbotic_relationships}\vspace{-2mm}
\end{figure}

Thus,  mutualistic or competitive symbiotic relationships can exist  \cite{Liang2022,Guo2019,Long2020}.  If primary and secondary data rates are $R_p$ and $R_s$, we can classify  SR systems as commensal SR (CSR) or parasitic SR (PSR)  \cite{Long2020} (Fig. \ref{fig:symbotic_relationships}). In CSR, $R_s \ll R_p,$ and the secondary signal is treated as a multi-path component \cite{Long2020}. 
In PSR, $R_p=R_s$ and secondary signals directly interfere with primary signals (parasitism). Both systems may also try to achieve the maximum transmission rate by competing resources and harming each other (interference).  Hence, multi-objective optimization is necessary to improve the performance of both primary and secondary simultaneously \cite{ Guo2019,  Zhang2021, Xu2015}. 

Backscatter communication (\bc)  networks are a natural candidate for symbiotic wireless network designs because they consist of tags, which can scavenge energy off of ambient or external radio frequency (RF) signals. Thus, tags are reminiscent of parasitical organisms.  These passive tracking tags are, by design, small and thin. Moreover, they are inexpensive, {i.e., \num{0.01}\$$\sim$\num{0.5}\$ per tag,} and do not require batteries (so they last for many years) \cite{ Diluka2022}. Hence,  tags eschew active RF components, resulting in ultra-low-power consumption, low complexity, and low cost, which are massive benefits \cite{Diluka2022, Rezaei2023}. {Energy harvesting (EH) can power tags, which offers the potential to save the cost of batteries and their replacement/recharging.}  Motivated by these advantages, the 3rd generation partnership project (3GPP)  is developing \bc standards to support logistics, warehousing, healthcare,   and many others \cite{Huawei, Passive_IoT_design}. 

We focus on  Ambient BackCom (\abc), where the tags reflect and/or absorb an ambient  RF signal from legacy (primary)  sources, such as cellular BSs, television (TV) towers, Wi-Fi access points, etc. \cite{rezaei2020large}. Therefore, \abc requires no new spectrum allocations, no licensing and regulatory hurdles,  and no need for dedicated RF emitters. Because of these massive advantages, the symbiosis of  \abc and conventional primary networks should yield the best of both worlds. Another advantage is that the primary network needs only minor modifications. For instance,    we can modify the primary BS to support both networks and improve their interactions. Moreover, primary users can decode both primary data and tag-emitted data via successive interference cancellation (SIC) decoding techniques \cite{Long2020}.

As the primary symbiont, we choose a non-orthogonal multiple access (NOMA)   network (Fig. 1)    \cite{Ding2017Application, Ding2017, Yongjun2021, Xu2015}. NOMA 
 allows some interference among multiple users in return for them to access the same spectrum resource simultaneously \cite{Ding2017Application, Ding2017}. 
 { NOMA  offers a range of advantages, including improved SE, enhanced user fairness, support for heterogeneous devices, and efficient handling of massive connectivity. {It can cater to the needs of diverse devices with varying quality of service requirements (i.e., essential service assurance in terms of metrics such as data rate, end-to-end delay, and loss, needed for specific applications {\cite{Xu2015}}) and channel conditions {\cite{Yongjun2021,Ding2017Application}}}. By accommodating these differences, NOMA strikes a delicate balance between system throughput and user fairness \cite{Ding2017}. One of the notable benefits of NOMA is its ability to support massive connectivity, a crucial requirement for numerous applications in today's wireless networks. This capability positions NOMA as a promising solution for meeting the demands of future communication networks, particularly in terms of ultra-low latency and ultra-high connectivity \cite{Chen2020NOMA, Zhang2019, Elsayed2022, Zhuang2022}. As a result of these advantages, the integration of NOMA into wireless systems has become a subject of intense research interest recently. The exploration of such integrated systems holds tremendous potential for enhancing the performance and capabilities of wireless communication networks \cite{Yongjun2021,Ding2017Application}. 
 Consequently, NOMA can serve as an ideal primary access scheme in the context of SR networks, allowing for significant improvements in overall SE and EE  \cite{Chen2020NOMA, Zhang2019, Elsayed2022, Zhuang2022}.}

\subsection{{Previous Contributions on NOMA-Assisted SR Systems}}
Symbiotic cellular  NOMA and \abc have been investigated   \cite{Khan2021, Xu2021NOMA, le2019outage, ahmed2021backscatter, Khan2021Energy}. These works are summarized in Table \ref{comparison}. They maximize the symbiosis benefits to the primary network without considering the tag's performance. In particular,  in  \cite{Khan2021}, a single-antenna BS serves two NOMA  primary users and a tag in a   CSR setup. {The symbiosis benefit to the primary system (rate increase) is maximized under 
  SIC decoding error by jointly optimizing power allocation factor, $\rho$,  and tag's reflection coefficient, $\alpha$.}  For the same system setup,  the work \cite{Xu2021NOMA} investigates the symbiosis benefits to the primary system in terms of EE with perfect SIC. Reference \cite{ahmed2021backscatter} also investigates the primary system's EE by extending  \cite{Khan2021} to a multi-cell scenario. Reference \cite{le2019outage} evaluates the outage probability of primary users in a PSR setup with a multi-antenna BS.

Symbiotic NOMA-assisted vehicular network with tag is investigated \cite{Khan2021Energy}.  Roadside units transmit information to vehicles through NOMA, and tags assist this task in a  CSR setup. This work maximizes the primary system's EE subject to rate constraint under imperfect SIC. Reference \cite{Khan2021Rate} considers a similar scenario where the roadside units relay information from the BS to the vehicles, and that the tags reflect their data to the vehicles using the relayed signals. This is a PSR network, and the BS power allocation and the roadside units can be optimized to maximize the minimum achievable rate of the vehicles. 

Unlike  \cite{Khan2021, Xu2021NOMA, ahmed2021backscatter, Khan2021Energy},   \cite{Chen2020NOMA, Zhang2019, Elsayed2022, Zhuang2022} study the symbiosis benefits/harms to both the tag and the primary users. In particular, \cite{Chen2020NOMA} considers two downlink NOMA users.  The symbiosis benefits to the tag (the rate) are then maximized under a power consumption constraint while preserving the primary outage by jointly optimizing $\alpha$ and BS transmit power, $p_t.$ For the same network, the outage probability and ergodic rates of the primary users and tag are investigated in \cite{Zhang2019} and \cite{Elsayed2022}  for  Rayleigh and Nakagami-m fading, respectively.  Reference \cite{Zhuang2022} considers a RIS-assisted NOMA system. Here, the  RIS assists the primary cell-edge user's transmission assuming direct communication with the transmitter is unavailable. The symbiosis benefit in terms of the overall EE of the system is maximized by jointly optimizing the RIS phase shifts and the power allocation for the NOMA users, considering their rates.  

These works do not evaluate how beamforming designs add to symbiosis. Thus,  the full benefits and harms of symbiosis should be further investigated.

\begin{table*}[!t]
\caption{Summary of Related Works.}
\label{comparison}
 \begin{center}
\begin{threeparttable}
\scalebox{1}{
\begin{tabular}{|l|l|l|lll|l|l|l|}
\hline
\multirow{2}{*}{Setup} & \multirow{2}{*}{Reference} 		& \multirow{2}{*}{Objective} 	& \multicolumn{3}{l|}{\quad \quad \quad Constraints}                                                                              & \multirow{2}{*}{Variables$^\dagger$}         								&\multirow{2}{*}{Methodology} & \multirow{2}{*}{\begin{tabular}[c]{@{}l@{}}Imperfect\\ SIC\end{tabular}}	\\ \cline{4-6}
                       &                            		&                            	& \multicolumn{1}{l|}{\begin{tabular}[c]{@{}l@{}}Primary\\ rate\end{tabular}} & \multicolumn{1}{l|}{\begin{tabular}[c]{@{}l@{}}Tag\\ rate\end{tabular}} & EH &                              												&								&	\\ \hline
\multirow{4}{*}{CSR}   & \cite{Khan2021}                    & Primary sum-rate   		 	& \multicolumn{1}{l|}{--}                                                      & \multicolumn{1}{l|}{--}       & --   &  \begin{tabular}[c]{@{}l@{}}$\rho$, $\alpha$\end{tabular}      	        & Sub-gradient &				\multicolumn{1}{l|}{\cmark}	    \\ \cline{2-9} 
                       & \cite{Xu2021NOMA}    			    & Primary EE 		    		& \multicolumn{1}{l|}{\cmark}                                                      & \multicolumn{1}{l|}{\xmark}       &  \xmark  &  \begin{tabular}[c]{@{}l@{}}$\rho$, $\alpha$\end{tabular}      	        & Dinkelbach's &	\multicolumn{1}{l|}{\xmark}				    \\ \cline{2-9} 
                       & \cite{ahmed2021backscatter}  	    & Primary EE 		    		& \multicolumn{1}{l|}{\cmark}                                                      & \multicolumn{1}{l|}{\xmark}       &  \xmark  &  \begin{tabular}[c]{@{}l@{}}$\rho$, $\alpha$\end{tabular}  		        & Dinkelbach’s & \multicolumn{1}{l|}{\cmark} 				    \\ \cline{2-9} 
                       & \cite{Khan2021Energy}  		    & Primary EE  		    		& \multicolumn{1}{l|}{\cmark}                                                      & \multicolumn{1}{l|}{\xmark}       &  \xmark  &  \begin{tabular}[c]{@{}l@{}}$\rho$, $\alpha$\end{tabular}  		        & Dinkelbach’s& \multicolumn{1}{l|}{\cmark} 					    \\ \hline
\multirow{5}{*}{PSR (parasitism)}   & \cite{Chen2020NOMA}   		        & Tag rate      		    		& \multicolumn{1}{l|}{\cmark}                                                      & \multicolumn{1}{l|}{\cmark}       &  \cmark  &  \begin{tabular}[c]{@{}l@{}}$\rho$, $\alpha$\end{tabular}  		        & --  & \multicolumn{1}{l|}{\xmark}							    \\ \cline{2-9} 
                       & \cite{Zhuang2022}  			    & EE  				    		& \multicolumn{1}{l|}{\cmark}                                                      & \multicolumn{1}{l|}{\cmark}       & \xmark   &  \begin{tabular}[c]{@{}l@{}}$\rho$, $\boldsymbol{\Theta}$\end{tabular}         & Dinkelbach’s &\multicolumn{1}{l|}{\xmark} 				    \\ \cline{2-9} 
                       & \cite{Khan2021Rate}   		        & Max-min rate 		    		& \multicolumn{1}{l|}{--}                                                      & \multicolumn{1}{l|}{--}       & --   &  \begin{tabular}[c]{@{}l@{}}$\rho$\end{tabular} 					        & Max-min \& Sub-gradient & \multicolumn{1}{l|}{\xmark}  	    \\ \cline{2-9} 
  & \multirow{2}{*}{\textbf{This Paper}}        & WSR    			& \multicolumn{1}{l|}{\cmark}                                                      & \multicolumn{1}{l|}{\cmark}       & \cmark &  \begin{tabular}[c]{@{}l@{}}$\mathbf{w}$, $\rho$\end{tabular}             & AO \& FP & \multicolumn{1}{l|}{\cmark}      \\ \cline{3-9} 
                       &                          			& Transmit power       			& \multicolumn{1}{l|}{\cmark}                                                      & \multicolumn{1}{l|}{\cmark}       &  \cmark  &  \begin{tabular}[c]{@{}l@{}}$p_t$, $\mathbf{w}$, $\rho$\end{tabular}      & AO \& SDR  & \multicolumn{1}{l|}{\cmark}                        \\ \hline
\end{tabular}}
\begin{tablenotes}
 \scriptsize{\item[$\dagger$] The variables $p_t$, $\mathbf{w}$, $\rho$, $\alpha$, and $\boldsymbol{\Theta}$ denote transmit power, transmit beamforming, power allocation factor, reflection coefficient, and  RIS phase shifts.}
\end{tablenotes}
\end{threeparttable}
\end{center}
\end{table*}
\subsection{Problem Statement and Contributions}

In the paper, we investigate several fundamental questions about symbiotic NOMA (primary) +\abc (secondary) networks (Fig. \ref{fig:SystemFig}). 
These are (i) how can the BS beamforming design be improved to maximize the benefits or minimize the harms for both networks?, and (ii) how to ensure tag activation with enough energy?   
 To answer these critical questions, we consider a PSR network of multiple users and one passive tag (Fig.~\ref{fig:SystemFig}). The BS uses a common beamformer to shape its RF signal to serve users and tag to achieve symbiosis. The tag harvests energy from the BS signal and reflects it to send data to the users. The nearest user decodes the tag data via SIC. The SIC decoding process is imperfect. Thus, the main problems are  (a) determining how the BS should optimally transmit its signal while using its multiple antennas' spatial degrees of freedom, and (b)  how it should allocate the total power to different users and the tag to maximize the benefits of symbiosis. Another goal is for it to minimize the transmit power under the rate and energy constraints.

In addition, we aim to address the following gaps in the literature: 1) Minimizing power consumption, supporting the minimum rate targets, and ensuring the EH potential of tags are the key IoT requirements.   Digital or analog beamforming can minimize transmit power.   However, previous studies have yet to consider this issue.   2)  Weighted sum-rate maximization (WSRMax) incorporates users' and the tag's different rate requirements. However, no prior studies have done this nor considered imperfect SIC decoding. However,  error propagation occurs with imperfect  SIC. (3)  The RF signal sent by the BS also must "wake up" the tag.  The minimum activation power level is about  \qty{-20}{\dB m} for commercial tags   \cite{tags}.  Except for  \cite{Chen2020NOMA}, no other study has considered the EH requirement of the tag.  Table \ref{comparison} compares and contrasts our work with the relevant studies.

To the best of our knowledge, no work has investigated jointly maximizing the symbiosis benefits, i.e., overall rate and tag's EH increase, and minimizing the symbiosis harms, i.e., interference and primary rate reduction. Besides, no work considers the various rate requirements of different devices, the transmit power consumption, and  BS beamforming designs. However,  the BS can optimize its beamforming design to facilitate symbiosis more effectively and efficiently by suppressing the harms and enhancing the benefits. Optimal beamforming also reduces the power consumption for a given desired performance, making the symbiosis green (energy-efficient). We design the BS beamformer and power allocation factor to jointly improve the benefits and minimize the harm of imperfect SIC. Other than  \cite{Khan2021, ahmed2021backscatter, Khan2021Energy}, which mainly focuses on the primary rate, no studies have considered the impact of SIC on symbiosis. Furthermore, most relevant works \cite{Khan2021, Xu2021NOMA, Chen2020NOMA, Zhuang2022} consider the basic two-user NOMA setup. Thus, we aim to fill these fundamental gaps for an  SR network with $K (\ge 2)$ NOMA users and a tag.

The main contributions of this paper are  as follows:
\begin{enumerate}
    \item To achieve symbiosis, we maximize the WSR of NOMA users and the tag. The problem requires joint optimization of the digital/analog beamforming vector and the power allocation factor. Due to intricately coupled variables in the signal-to-interference-plus-noise ratio (SINR)/signal-to-noise ratio (SNR), this problem has a non-convex objective and constraints. Thus, common convex algorithms are not able to handle it.  Hence, we split it  into two sub-problems, namely  $\mathbf{P1}_{\rm{w}}$ \eqref{P1w_prob} and $\mathbf{P1}_{\rho}$ \eqref{P1r_prob}. We then employ alternating optimization (AO) and solve these iteratively using the fractional programming (FP) technique.
    
    \item  Another way to achieve symbiosis is to minimize the  BS transmit power under the given constraints. Transmit power minimization (TPMin) involves jointly optimizing the digital beamformer, transmit power, and power allocation factor. Again,  the constraints are not convex functions of the optimization variables, and this problem is non-convex. We thus separate it into two sub-problems, namely $\mathbf{P2}_{\rm{p}}$ \eqref{P2p_prob} and $\mathbf{P2}_{\rho}$ \eqref{P2r_prob},  and apply iterative  AO and semi-definite relaxation (SDR) algorithm. 
    
    \item  The proposed algorithms improve symbiotic performance gains.  Significantly, there is no additional processing or changes to the tag.  Thus, symbiotic gains and passive tags can help realize passive IoT networks. 
    
    \item Finally,  we present numerical examples to evaluate the symbiotic gains through the proposed solutions.
\end{enumerate}

\subsection{Structure and Notations}

The paper is organized as follows. Section \ref{system_modelA} introduces the system model, tag operations, and the achievable rates of the primary users and the tag. In Section \ref{Sec:Optimization_problem_formulation}, we formulate the WSRMax and  TPMin problems. We present the AO solutions in Section \ref{Sec:proposed_solution}. In Section \ref{sim}, simulation examples are presented for performance evaluations. Section \ref{conclusion} concludes the paper and outlines future research directions.

\textit{Notation}: Lower-case bold and upper-case bold denote vectors and matrices.  $\mathbf{I}_n$ is  the $n \times n$ identity matrix. $\mathbf{A}^\mathrm{T}$, $\mathbf{A}^\mathrm{H}$, $\text{Tr}(\mathbf{A})$, and $[\mathbf{A}]_{(m,n)}$ denote transpose,  Hermitian transpose, trace, and the $(m,n)$-th element of  $\mathbf{A}$.  $\mathbb{E} \{ x \}$ is the mean of $x$.  Moreover, $\mathcal{CN}(\bm{\mu},\mathbf R ) $ is a complex Gaussian  vector with  mean   $\bm \mu$ and co-variance   $\mathbf R$. Finally, $\mathcal{K} \triangleq \{1,\ldots,K\}$,  $\mathcal{K}_k \triangleq \{1,\ldots,k-1\}$, $\mathcal{K}_k' \triangleq \{k+1,\ldots,K\}$, and $\mathcal{K}_0 \triangleq \{0,1,\ldots,K\}$. The variables are defined in Table \ref{table_notations_param}.

\begin{table*} 
\caption{Notations} \label{table_notations_param}
\centering 
\begin{tabular}{|c l |c l|} 
\hline 
Notation & Definition & Notation & Definition \\ [0.2ex] 
 \hline 
$M$ &BS transmit antennas  & $U_k$ & $k$th user  \\ 
$K$ &Number of users  & $\xi_j$ & SIC quality   \\ 
$\mathbf{w}$ & BS beamforming vector & $p_t$ & BS transmit power\\
$\zeta_{\mathbf{a}}$ &Large scale path-loss of channel $\mathbf{a}$ &  $\mathbf{h}_k, \mathbf{f}_k, q_k$ & BS-to-$U_k$, BS-to-tag, tag-to-$U_k$ channels \\ %
 $\alpha$ & tag's reflection coefficient &  $c(n)$ & Tag's signal\\
$p_a$ &Tag's received power  & $p_h$ & Tag's harvested power \\ %
$\eta_b$ & EH circuit conversion efficiency & $\rho_k$& Power allocation coefficient for $U_k$\\
$s_k(n)$ & Signal of the $k$th user &$x(n)$ & Users' superimposed signal\\ [0.2ex] 

\hline 
\end{tabular}
\label{table:nonlin} 
\end{table*} 

\section{System Model and Preliminaries}\label{system_modelA}
Here,  we describe the system model, channel model, and transmission model in detail.

\vspace{-3mm}
\subsection{System and Channel Models}\label{Sec:system_model}
Fig.~\ref{fig:SystemFig} shows a symbiotic network that comprises a BS, as the primary transmitter, equipped with $M\geq 1$ antennas, $K$ primary users each with a single antenna, denoted by $U_k, k \in \mathcal{K}$, and a single-antenna tag, which are all located nearby, e.g., within a room. The  BS serves the primary NOMA users. {We assume that the tag is located in the proximity of the BS. The tag communicates by reflecting the incident RF signal from the BS to communicate the tag's data. Without loss of generality,   the user with the strongest channel gain, i.e., $U_1$, decodes the tag's data.}
Hence, the BS transmits its data to the users via transmit beamforming, simultaneously enabling the tag to send its data, which is decoded by $U_1$.

The  BS may use two beamforming designs.
\begin{enumerate}\label{beamforming}
    \item \textit{Digital Beamforming}: This requires a dedicated RF chain per antenna element,  and phases and amplitudes are digitally controlled by baseband processing. It provides the best beam control but is expensive from both cost and power consumption perspectives for large-scale antenna arrays \cite{Kasemir2009}.
    
    \item \textit{Analog (Constant-Modulus) Beamforming}: A single RF chain is connected to the antenna array and the beam is controlled by adjusting analog phase shifts.  Thus, the beamforming vector has constant-modulus elements. The number of RF chains should exceed the number of data streams. At the same time, the beamforming gain and diversity order are given by the number of antenna elements  \cite{Molisch2017}. Additionally, analog beamforming can significantly reduce power consumption and hardware complexity in large-scale antenna arrays compared to digital beamforming.
\end{enumerate}

For notational brevity, the set of BS  antennas is denoted as $\mathcal{M}=\{1,\ldots, M\}$. We consider a block,  flat-fading channel model. During each fading block, the channel coefficient vector between user  $U_k, k \in \mathcal{K}$ and the BS is given as $\mathbf{h}_k  \in \mathbb{C}^{M \times 1}$. Moreover, $ \mathbf{f} \in \mathbb{C}^{M \times 1}$ is the forward-link channel vector between the tag and the BS, and $q_k \in \mathbb{C}$ is the channel coefficient between the tag and $U_k$.  All these channels  can be   represented in a unified manner     as
\begin{eqnarray}\label{pathloss}
    \mathbf{a} = \zeta_{\mathbf{a}}^{1/2} \tilde{\mathbf{a}},
\end{eqnarray}
where $\mathbf{a} \in \{ \mathbf{h}_{k},\mathbf{f}, {q}_{k} \}$, $k \in \mathcal{K}$ and  ${\zeta}_{\mathbf{a}}$ accounts for the large-scale  path-loss and shadowing. Moreover,   $\tilde{\mathbf{a}} \sim  \mathcal{CN}\left(\mathbf{0},\mathbf{I}_{M} \right)$ captures  Rayleigh fading. Note that $q_k = \zeta_{{q}_k}^{1/2} \tilde{{q}}_k$ and $\tilde{{q}}_k \sim  \mathcal{CN}\left({0},1 \right)$. We will use this unified representation throughout the paper for simplicity. 

{The SR system (Fig.~\ref{fig:SystemFig})  operates in the time-division-duplexing (TDD) mode, where the same frequency band is used for both uplink (UL)  and downlink (DL)  transmissions. Consequently, the radio channel is reciprocal because it has the same characteristics in both UL and DL directions.  Thus, the BS  transmitter can estimate the DL  channel from the sounding on the UL channel. }  Alternatively, in the frequency-division-duplexing (FDD) mode, UL and DL channels occupy different frequency bands.  However, TDD offers higher SE and the ease of channel state information (CSI)  estimation using  UL pilots over two training phases  \cite{Long2020}.   For example,  each NOMA user transmits two sets of pilots. During the first phase, the tag goes into complete signal absorption with reflections. This allows the BS to estimate the direct-link channel, $\mathbf{h}_k$. In the second phase, the tag switches its impedance into a fixed value to represent a pre-determined symbol, $c_0$, and fixed reflection coefficient, $0 < \alpha <1.$ Hence, the BS can estimate the composite channel, $\mathbf{h}_k+ \sqrt{\alpha} c_0 q_k \mathbf{f} $, using the training pilots.  Thus,  the BS can obtain the backscatter channel, $ q_k \mathbf{f}, k \in \mathcal{K}$, by subtracting the estimated direct-link channel component from the estimated composite channel. The  BS can then estimate $\mathbf{f}$ through the expectation-maximization algorithm \cite{Ma2018}. {Based on the TDD model and the existence of reliable channel estimation methods, it is customary to assume perfect CSI availability. This assumption is then leveraged for various tasks such as DL resource allocation, beamforming, power allocation, and data decoding at the users. It is important to note that this assumption is extensively employed in the majority of research studies \cite{Guo2019, Long2020, Zhang2021}.}

\subsection{Data Transmission and EH at the Tag}\label{passive_tag}
The tag must perform two operations, namely  EH and data transmission. These operations can be done in one of the two modes: (i) time-switching, where the two operations are done in two distinct time slots, and (ii) power-splitting, where both operations occur simultaneously, but each captures a fraction of the incident RF power $p_a$ \cite{Diluka2022}. In words,  the tag reflects   $\alpha p_a$  and harvests $p_l = (1-\alpha)p_{a}.$  To perform data transmission, the  tag selects  symbol   $c(n)$ $(  \mathbb{E}\{\vert c(n) \vert^2\} = 1)$, from a  $\bar{M}$-ary constant-envelope modulation. For more details on the tag modulation process, please see  \cite{Rezaei2023} and references therein. Thus, we consider the power-splitting mode, which is widely used for passive tags  \cite{Diluka2022}.

We assume a  linear EH model for its tractability. It predicts the harvested power as   $p_h=\eta_b p_l$, where $\eta_b \in (0,1]$ is the power conversion efficiency. Although practical EH circuits exhibit non-linear characteristics,  the linear model can be accurate for certain operating regions \cite{Wang2020}. Since the tag is batteryless, it must harvest more than a  minimum threshold ($p_b$) to maintain its operation, i.e., $p_h \ge p_b$. This threshold is about \qty{-20}{\dB m} for commercial passive tags \cite{Zhang2021, tags}.

\subsection{Transmission Model}
The BS transmits  signal $s_k(n)$ to $U_k$ ($k \in \mathcal{K}$) at the $n$-th time slot, where $E\{|s_k(n)|^2\}=1$, with transmission power $\rho_k p_t$, where  $p_t$ is the total BS power  and $\rho_k \in (0,1)$ is the power allocation coefficient for  $U_k$. With $K$-user NOMA, $s_k(n), k \in \mathcal{K}$ are superimposed as
\begin{eqnarray}\label{transmit_signal}
    x(n) =  \sum\nolimits_{k\in \mathcal{K}}{\sqrt{\rho_k p_t} s_k(n)},
\end{eqnarray}
where the set of  power coefficients $\{\rho_{k}\}$ satisfies $\sum_{k\in \mathcal{K}}{\rho_{k}} = 1$. Furthermore,  the different  data signals are mutually uncorrelated; $\mathbb{E}\{s_{k}(n) s_{i}^*(n) \}= \delta(k-i)$,  where $k,i \in \mathcal{K}$ \cite{Rezaei2020NOMA, Diluka2020}. Therefore, $\mathbb{E}\{ |x(n)|^2\} =  \sum_{k\in \mathcal{K}} {\rho_k p_{t}} = p_{t}$. 

The tag  harvests energy from  $x(n)$  and  also reflects $x(n)$ its binary signal, $c(n)$, $\mathbb{E}\{|c(n)|^2\} =1$. The users thus receive the tag reflection and $x(n)$. Here, we assume that the propagation delay differences are negligible \cite{Xu2021NOMA}. Therefore, the received signal at $U_k, k \in \mathcal{K}$ is given as 
\begin{eqnarray}\label{transmissin_model}
   y_k(n) = \mathbf{h}_k^{\rm{H}} \mathbf{w} x(n) + \sqrt{\alpha}  \mathbf{g}_k^{\rm{H}} \mathbf{w} x(n) c(n)+n_k(n),
\end{eqnarray}
where the first term in $y_k(n)$ is the direct-link signal and the second term is the backscatter-link signal. Besides, $\mathbf{g}_k = \mathbf{f} q_k$, $k \in\mathcal{K}$, is the backscatter link between $U_k$ and the BS. Moreover, $\mathbf{w} \in \mathbb{C}^{M \times 1}$ denotes  the BS  beamforming vector, where (i) $\Vert\mathbf{w} \Vert^2 = 1$ for digital beamforming and (ii) $\vert[\mathbf{w}]_m \vert = \frac{1}{\sqrt{M}} \, \forall m \in \mathcal{M}$ for constant-modulus beamforming (Section~\ref{beamforming}). Furthermore, $n_k \sim \mathcal{CN}(0,\sigma^2)$ denotes the additive white Gaussian noise (AWGN) at $U_k$. In this paper, we assume that the primary signal $s_k(n), k \in \mathcal{K}$ and the tag signal $c(n)$ are standard circular symmetric complex Gaussian variables,  $\mathcal{CN}(0,1)$ \cite{Long2020, Chu2020}.

{Without loss of generality, we assume that   $U_1,$  which has the largest channel gain, decodes the tag signal.} Hence, to  guarantee successful decoding,  the  users are ordered based on the effective channel gains:
\begin{eqnarray}\label{NOMA_c}
    \vert \mathbf{h}_1^{\rm{H}} \mathbf{w} \vert^2 \ge \vert \mathbf{h}_2^{\rm{H}} \mathbf{w} \vert^2 \ge \ldots \ge \vert \mathbf{h}_K^{\rm{H}} \mathbf{w} \vert^2.
\end{eqnarray}

{Note that the primary network uses power-domain NOMA. The tag is a secondary device that simply reflects a part of the BS signal,  i.e, $\alpha \vert \mathbf{f}^{\rm{H}} \mathbf{w} \vert ^2 p_t$, received by the users as $\alpha \vert (\mathbf{f}q_k)^{\rm{H}} \mathbf{w} \vert ^2 p_t$ for $k \in {\mathcal{K}}$. The backscattered signal from the tag is typically weaker than the direct-link signal because of the power loss during reflection at the tag and the double path losses originating from the cascaded link  $\mathbf{f}q_k$ \cite{Diluka2022, Long2020, Chen2020NOMA}. Hence, \eqref{NOMA_c} always ensures SIC decoding requirements for the primary network. Besides, at each user, the tag signal is treated as interference when decoding the user
signals. The users thus require no prior information about the tag signal.}

Therefore, higher power is allocated to the user with lower effective channel strength, i.e., $\rho_{1}  \leq \cdots \leq \rho_{K}$ where $U_k$ applies SIC to decode $s_k(n)$. More precisely, $U_k$ decodes $s_k(n)$ of  the users with higher powers $( i \in \mathcal{K}_k')$, and then subtracts them  from the received signal ${y}_{k}(n)$ \eqref{transmissin_model}. Moreover, $U_k$ treats the signals of the other users with higher effective channel strength ($ i \in \mathcal{K}_k$),   and the tag's signal as interference. User 
 $U_1$ performs SIC to decode $s_1(n)$ and other users' signals and subtracts them from the received signal when decoding the tag's signal.

 
Mathematically, the condition that $U_k, k \in \mathcal{K}$ can perform SIC and decode the data intended for the users with lower effective channel strength, can be written as \cite{Hanif2016, Rezaei2020NOMA} 
\begin{align}\label{condition1}
    \mathbb{E}\{\text{log}_2(1+\gamma_{k}^{i})\} \geq  \mathbb{E}\{\text{log}_2(1+\gamma_{i}^{i})\}, \quad  i \in \mathcal{K}_k'. 
 \end{align}
{In \eqref{condition1}, $\gamma_{k}^{i}$ is the effective SINR of  $U_i$ at  $U_k$, when $U_k$ decodes the signal intended for $U_i$ with the desired power $\rho_i p_t \vert \mathbf{h}_k^{\rm{H}} \mathbf{w} \vert^2$, treating the signals of $U_j, j \in \mathcal{K}_j,$ and the tag's signal as interference.} 
 
We must note that allocating higher powers to users with lower effective channel strength also yields a non-trivial data rate for them \cite{Hanif2016}.

SIC  decoding errors may occur and   $U_k, k \in \mathcal{K}$ may not always perfectly cancel the interference, causing error propagation \cite{ahmed2021backscatter, Khan2021}. Thus, the received signal at $U_k$ {\eqref{transmissin_model}} after an imperfect SIC process can be written as {\mbox{\cite{Li2019, Chen2019}}}
\begin{align}\label{Eqn:sic}
\nonumber {\bar y}_{k}(n) &= y_{k}(n)- \mathbf{h}_k^{\rm{H}} \mathbf{w} \sum\nolimits_{j \in \mathcal{K}_k'} \sqrt{\rho_j p_t} \hat{s}_j(n)\\
 \nonumber &=\underbrace{ \mathbf{h}_k^{\rm{H}} \mathbf{w} \sqrt{\rho_k p_t}{s}_k(n)}_{T_0: \text{desired signal}} + \underbrace{\mathbf{h}_k^{\rm{H}} \mathbf{w} \sum\nolimits_{j \in \mathcal{K}_k} \sqrt{\rho_j p_t}{s}_j(n)}_{T_1: \text{interference after SIC}}  \\ 
 \nonumber &+ \underbrace{\mathbf{h}_k^{\rm{H}} \mathbf{w} \sum\nolimits_{j \in \mathcal{K}_k'} \sqrt{\rho_j p_t}\left({s}_j(n) -  \hat{s}_j(n) \right)}_{T_2: \text{{error propagation due to imperfect SIC}}} \\
 &+ \underbrace{\sqrt{\alpha}  \mathbf{g}_k^{\rm{H}} \mathbf{w} x(n) c(n)}_{T_3: \text{interference from the tag}} + n_k(n),
\end{align}
where $T_0$ is the desired signal, $T_1$ is the interference caused by the signals of the users which are considered as the interference at $U_k$, and  {$T_2$  represents the error propagation due to the imperfect SIC.} In \eqref{Eqn:sic}, $\hat{s}_{j}(n)$ is the decoded signal (the estimate of  ${s}_{j}(n)$) of $U_j$  by $U_k$. 

{In practice, the residual interference caused by imperfect SIC is a complicated function of multiple factors, e.g., coding/modulation-related parameters, channel-related issues (fading and shadowing), etc {\cite{Chen2019}}. However,  a linear function can effectively represent the SIC behavior and the relationship between the residual interference and the power of the received signal {\cite{Chen2019, Chen2018, Li2011}}.}
Without loss of generality, leveraging  linear minimum mean square error (MMSE) estimation, $s_{j}(n)$  and its estimate $\hat{s}_{j(n)}$ can be assumed as jointly Gaussian with a certain  correlation coefficient \cite{Rezaei2020NOMA, Li2019}, i.e.,
\begin{equation}\label{Eqn:decod}
s_{j}(n) = \xi_{j} \hat{s}_{j}(n) + e_{j}(n),
\end{equation}
where $\hat{s}_{j}(n)\!\sim \!\mathcal{CN}({0},1) $, $e_{j}(n) \!\sim\! \mathcal{CN}({0},\sigma^2_{e_{j}}/[1\!+\!\sigma^2_{e_{j}}])$ is the estimation error, statistically independent of $\hat{s}_{j}(n)$, and  $ \xi_{j} =1 /\! \sqrt{1\!+\!\sigma^2_{e_{j}}}$. {The correlation coefficient $\xi_{j} \in [0,1]$  shows the quality of the SIC-based decoding process, i.e., SIC quality, which refers to the accuracy of the SIC decoding process and characterizes the severity of the SIC imperfection when a user decodes the signal of another user}. The value of $\xi_{j}$ is determined by channel-related issues  (fading and shadowing) and other factors \cite{Li2011}; the greater its value, the greater the association between $\hat{s}_{j}(n)$ and $s_{j}(n)$ and better the SIC performance.

While practical limitations in SIC  often restrict many NOMA  systems to only two users, it is crucial to emphasize that there is no inherent theoretical limitation on the number of users, even in the presence of imperfect SIC. It is important to note that this paper specifically introduces the theoretical framework for the NOMA+SR system, which extends the boundaries of traditional power-domain  (i.e., non-symbiotic) NOMA systems without backscatter tags. 
By considering this framework, we can explore the potential for accommodating more users and overcoming practical restrictions associated with SIC, thus paving the way for enhanced NOMA system designs and performance. Hence, \eqref{Eqn:sic} expresses the theoretical post-processed signal of $U_k$ under imperfect SIC.

\subsection{Achievable Rate}

This section derives the user and the tag rates. Applying the rate inequality given in \eqref{condition1}, the rate of $U_k, k \in \mathcal{K}$ can be computed as
\begin{align}\label{Eqn:rate}
  \mathcal{R}_{k} =  {\rm{log}}_2(1+{{\gamma}}_{k}),
 \end{align}
 where ${{\gamma}}_{k}\triangleq \min(\gamma_{k}^{k},\gamma_{i}^{k}),  i \in \mathcal{K}_k$ to  guarantee that $U_i$ ($ i \in \mathcal{K}_k$) can perform SIC and decode the data of $U_k$ \cite{Hanif2016}. Following the same principle as \eqref{Eqn:sic}, for the received signal at the typical user $U_i, i \in \mathcal{K}$, by considering the first term in \eqref{Eqn:sic} as the desired signal and the remaining terms as an effective noise, the SINR of $U_k$ at $U_i$, $\gamma_{i}^{k}$, can be derived as \eqref{Eqn:rate_u2_min}\footnote{The distribution of $c(n) s_k(n)$ is approximated as circular symmetric complex Gaussian distribution which yields a lower bound of the achievable rate \cite{Long2020, Chu2020}.},
 \begin{figure*}
\begin{eqnarray}\label{Eqn:rate_u2_min}
    \gamma_{i}^{k} = \frac{\rho_k p_t \vert \mathbf{h}_i^{\rm{H}} \mathbf{w} \vert^2}{\vert \mathbf{h}_i^{\rm{H}} \mathbf{w} \vert^2 p_t \sum\nolimits_{j \in \mathcal{K}_k} {\rho_j }  + \vert \mathbf{h}_i^{\rm{H}} \mathbf{w} \vert^2 p_t \sum\nolimits_{j \in \mathcal{K}_k'} {\rho_j \left( 2-2 \xi_j \right)} +\alpha p_t  \vert \mathbf{g}_i^{\rm{H}} \mathbf{w} \vert^2 + \sigma^2}
 \end{eqnarray}
 \hrulefill
 \vspace{-4mm}
 \end{figure*}
where using \eqref{Eqn:decod}, $ \mathbb{E}\{\vert s_j(n)- \hat{s_j}(n) \vert^2 \} = 2 -2 \xi_j$.

On the other hand, $U_1$ decodes the tag's signal by performing SIC and subtracting its decoded signal and other users' signals. Hence, using \eqref{transmissin_model}, the received signal to decode the tag's signal after the imperfect SIC process can be written as
 \begin{eqnarray}\label{Eqn:sic_for_tag}
\nonumber {\bar y}_{1,b} &= & y_{1}- \mathbf{h}_1^{\rm{H}} \mathbf{w} \hat{x}(n) \\
&= & \mathbf{g}_1^{\rm{H}} \mathbf{w} x(n)c(n) + \mathbf{h}_1^{\rm{H}} \mathbf{w}\left(x(n) - \hat{x}(n)\right) + n_1(n),
\end{eqnarray}
where $\hat{x}(n) = \sum_{j \in \mathcal{K}}\sqrt{\rho_j p_t} \hat{s}_j(n)$ and $ x(n) - \hat{x}(n) = \sum_{j \in \mathcal{K}} \sqrt{\rho_j p_t}({s}_j(n) -  \hat{s}_j(n))$.

Therefore, the rate of the tag after imperfect SIC is given as
\begin{eqnarray}\label{Rate_BD}
    {R}_{0} = \mathbb{E}_x \left \{{\rm{log}}_2(1+\gamma_{0}^b) \right \},
\end{eqnarray}
where\footnote{To simplify the analysis, we substitute the $\vert s_j(n)- \hat{s_j}(n) \vert^2$ in $\gamma_0^b$'s denominator with its average value, $ \mathbb{E}\{\vert s_j(n)- \hat{s_j}(n) \vert^2 \} = 2 -2 \xi_j$, yielding a lower bound.} 
\begin{eqnarray}\label{Eqn:snr_BD}
   \gamma_{0}^b = \frac{\alpha \vert \mathbf{g}_1^{\rm{H}} \mathbf{w} \vert^2 \vert x(n) \vert^2 }{\vert \mathbf{h}_1^{\rm{H}} \mathbf{w}  \vert^2  \sum \nolimits_{j \in \mathcal{K}}{{\rho_j p_t} \left( 2-2 \xi_j \right)}  +\sigma^2}.
\end{eqnarray} 
In \eqref{Rate_BD}, $\gamma_{0}^b$ is the received SINR of the tag at $U_1$. By taking the average over $x(n)$ in \eqref{Rate_BD}\footnote{For  $s_i(n) \sim \mathcal{CN}(0,1)$, the square envelope of $x(n)$ follows exponential distribution.}, the rate of the tag is given as 
\begin{eqnarray}\label{Eqn:rate_BD}
     R_{0} = -e^{\frac{1}{\gamma_{0}'}} {\rm{E}}_i\left(-\frac{1}{\gamma_{0}'}\right) {\rm{log}}_2 (e),
\end{eqnarray}
where $\gamma_{0}'$ is the average received SINR of the backscatter link and is given as
\begin{eqnarray}\label{Eqn:average_snr_BD}
   \gamma_{0}' = \frac{\alpha p_t \vert q_1 \vert^2 \vert \mathbf{f}_1^{\rm{H}} \mathbf{w} \vert^2}{\vert \mathbf{h}_1^{\rm{H}} \mathbf{w}  \vert^2  \sum \nolimits_{j \in \mathcal{K}}{{\rho_j p_t} \left( 2-2 \xi_j \right)} +\sigma^2}.
\end{eqnarray} 
Additionally, ${\rm{E}}_i(x) = \int_{-\infty}^{x} u^{-1} e^u du$ and it is worth noting that, $-e^{\frac{1}{\gamma_{0}'}} {\rm{E}}_i (-{1}/{\gamma_{0}'})$ is monotonically increasing and concave function of $\gamma_{0}'$ \cite{Long2020}.

As the tag must exceed the activation threshold,   Section \ref{passive_tag},  we get  the constraint 
\begin{eqnarray}
    \eta_b (1-\alpha)  p_a \ge p_b,
\end{eqnarray}
where $p_a = \vert \mathbf{f}^{\rm{H}} \mathbf{w} \vert ^2 p_t$ is the 
 tag's received power.

\section{Optimization Problem Formulation}\label{Sec:Optimization_problem_formulation}
This problem involves optimizing the   BS beamforming vector and transmit power, $\mathbf{w}$ and $p_t,$ and the    NOMA power allocation factors, $\boldsymbol{\rho} = [\rho_1, \ldots, \rho_K]$.  The optimization problem is to maximize the weighted network sum rate or minimize the BS transmit power. The rationale for these metrics is discussed below. Another goal is to ensure that all nodes achieve a minimum rate and that the tag meets the minimum energy threshold. To incorporate all these factors, we formulate two optimization problems: 1) The WSRMax problem to optimize the  BS digital/analog beamforming vector and the power allocation factor, and 2) The TPMin problem to optimize the  BS digital beamforming vector and power allocation factor. {Since analog beamforming keeps the beam amplitude constant, it is not effective for minimizing the transmit power. Hence, the TPMin problem considers digital beamforming only.}

{In the WSRMax approach, our objective is to maximize the overall data rate while maintaining fair and efficient spectrum utilization. To achieve this, we assign different weights to users and tags based on their priority or importance. By doing so, we ensure that the system meets the required rate targets while operating within the given BS transmit power  \cite{Yongjun2021, weeraddana2012weighted}. This approach provides a comprehensive solution that balances the needs of different users and optimizes the utilization of available resources. In contrast, TPMin focuses on supporting large-scale connectivity by minimizing the BS transmit power. By doing that, we can allocate the saved power to connect more devices, thereby enhancing the network's capacity for accommodating a higher number of users. This approach holds significant value in green/sustainable communication networks, where the emphasis lies on reducing energy consumption. By conserving energy, these networks enable extended network lifetimes, cost savings, and improved resource efficiency \cite{Toro2022}.}

{We assume a tag with a  fixed reflection coefficient, aligning with passive tags that possess limited processing capabilities and stringent energy constraints  \cite{Chen2020NOMA, Diluka2022}. Considering variable  $\alpha$ would increase the tag architectural complexity, cost, and form factor, which are counterproductive to the goals of passive IoT \cite{Huawei, Passive_IoT_design, Diluka2022}. Nonetheless, variable-$\alpha$ tags can help improve signal quality and enhance overall system performance. Thus,  the tag can optimize its performance based on factors such as the distance to the receiver, channel fading, and interference \cite{Xu2021NOMA, ahmed2021backscatter}. However, incorporating variable $\alpha$ into our resource allocation problems is an interesting topic that we plan to address in our future works.}

\vspace{-2mm}

\subsection{Weighted Sum-Rate Maximization}
WSRMax for an arbitrary set of interfering links plays a central role in many network control and optimization methods, as it can achieve different trade-offs between sum-rate performance and user fairness \cite{weeraddana2012weighted}. It provides a framework to integrate the rate constraints, the tag EH constraint, and the BS power constraint in order to simultaneously satisfy the demands of both networks and improve their performance symbiotically.  However, the power allocation entangles the beamforming design due to the rate constraints, complicating the problem. 

The WSRMax problem can be formulated as
\begin{subequations}\label{P1_prob}
    \begin{eqnarray}
        \mathbf{P1}:
        \underset {\mathbf{w},\boldsymbol{\rho}}{\rm{maximize}} && \sum\nolimits_{k\in \mathcal{K}_{0}} a_k R_k, \label{P1_obj} \\
        \text{subject to} && R_k \ge R_k^{\rm{th}},  \quad \text{for} \quad k \in \mathcal{K}_{0}, \label{P1_rateUi}\\
        && \eta_b (1-\alpha)  \vert \mathbf{f}^{\rm{H}} \mathbf{w} \vert ^2 p_t \ge p_b,  \label{P1_BD_EH}\\
        && \rho_{1} \leq \cdots \leq \rho_{K}, \quad \text{and} \quad \sum\nolimits_{k\in \mathcal{K}}{\rho_{k}} = 1, \label{P1_power} \qquad\\
        && \begin{cases}
      \Vert\mathbf{w}\Vert^2=1, &  \text{Digital,} \\ 
        \vert[\mathbf{w}]_m \vert = \frac{1}{\sqrt{M}}, m \in \mathcal{M} &\text{Analog}, \\ 
 \end{cases} \label{P1_preco}\\
        && \vert \mathbf{h}_1^{\rm{H}} \mathbf{w} \vert^2 \ge  \ldots \ge \vert \mathbf{h}_K^{\rm{H}} \mathbf{w} \vert^2,\quad \label{NOMA_constr}
    \end{eqnarray}
\end{subequations}
where  $a_k \in [0,1]$ for $k \in \mathcal{K}_{0}$ is the weight factor and $\sum_ {k \in \mathcal{K}_{0}}{a_k} =1$. Note that \eqref{P1_rateUi} constraint guarantees the required rate for $U_k, k \in \mathcal{K}$, and the tag, in which $R_k^{\rm{th}}, k \in \mathcal{K}_{0}$ denotes the minimal rate requirement of respective users and the tag, and $R_k$, $k \in \mathcal{K}_{0}$ is given in \eqref{Eqn:rate} and \eqref{Rate_BD}, respectively. Besides, \eqref{P1_BD_EH} is the minimum tag power requirement, \eqref{P1_power} is the transmit power allocation constraint at the BS, and  \eqref{P1_preco}  is the normalization constraint for the BS transmit beamforming, i.e., if the BS employs a digital beamformer, $\Vert\mathbf{w} \Vert^2=1$, or else if the BS uses constant-modulus beamforming, $\vert[\mathbf{w}]_m \vert = \frac{1}{\sqrt{M}}$ for $m\in \mathcal{M}$.  {In addition, the constraints  {\eqref{P1_power}} and {\eqref{NOMA_constr}} impose the necessary conditions to implement SIC decoding at the users {\cite{Hanif2016, Chen2019}}.}

\vspace{-2mm}
\subsection{Transmit Power Minimization}
Herein,  we aim to minimize the BS transmit power given primary users and tag rate requirements and the tag's EH constraint. To do that, we jointly optimize the transmit beamforming vector $\mathbf{w}$  and the power allocation factor $\boldsymbol{\rho}$. The resulting TPMin problem is given as 
\begin{subequations}\label{P2_prob}
    \begin{eqnarray}
        \mathbf{P2}:
        \underset {\mathbf{w},p_t,\boldsymbol{\rho}}{\rm{minimize}} && p_t, \\
        \text{subject to} && 
        R_k \ge R_k^{\rm{th}},   \quad \text{for} \quad k \in \mathcal{K}_{0},\label{P2_rateUi}\\ 
        &&\eta_b (1-\alpha)  \vert \mathbf{f}^{\rm{H}} \mathbf{w} \vert ^2 p_t \ge p_b,  \label{P2_BD_EH}\\
        && \rho_{1} \leq \cdots \leq \rho_{K}, \quad \text{and} \quad \sum\nolimits_{k\in \mathcal{K}}{\rho_{k}} = 1, \label{P2_power} \qquad\\
        &&\Vert \mathbf{w}\Vert^2=1,  \label{P2_preco} \\
        && \vert \mathbf{h}_1^{\rm{H}} \mathbf{w} \vert^2 \ge  \ldots \ge \vert \mathbf{h}_K^{\rm{H}} \mathbf{w} \vert^2. \quad
    \end{eqnarray}
\end{subequations}
We note that $\mathbf{P1}$ and $\mathbf{P2}$ are not convex since the objective function of $\mathbf{P1}$ and the corresponding constraints of $\mathbf{P1}$ and $\mathbf{P2}$ are not convex functions in either of the optimization variables, i.e., $\mathbf{w}$, $\boldsymbol{\rho}$, and $p_t$. Hence, we first transform these problems into convex forms and use the  AO technique to solve  $\mathbf{P1}$ and $\mathbf{P2}$. In what follows, we propose solutions to the above optimization problems.  

\section{Proposed Solutions}\label{Sec:proposed_solution}
First, we replace the rate of the tag, given in \eqref{Rate_BD}, with its lower bound for mathematical simplicity and to solve the proposed optimization problems as \cite{Zhang2014}  
\begin{eqnarray} \label{Eqn:upper}
    R_{0} \ge  R_{0}^{\rm{lb}} \triangleq \log_2 \left(1+ \left [\mathbb{E} \left \{ {1}/{\gamma_{0}^b} \right  \} \right ]^{-1} \right) \overset{(a)}{\mathop{=}} \, {\rm{log}}_2(1+\gamma_{0}),\quad
\end{eqnarray}
where $\gamma_0= \gamma_0'/2$, given in \eqref{Eqn:average_snr_BD}, and $(a)$ is obtained by using $\mathbb{E}  \{ {1}/{\gamma_{0}^b}   \} = {1}/{\mathbb{E}  \{ {\gamma_{0}^b}  \}} + {\sigma^2_{\gamma_{0}^b}}/{ [ \mathbb{E} \{ {\gamma_{0}^b}  \} ]^3}$, in which  $\gamma_{0}^b$ is given in \eqref{Eqn:snr_BD}.
We use the AO technique to obtain sub-optimal solutions for $\mathbf{P1}$ and $\mathbf{P2}$.

\subsection{Weighted Sum-Rate Maximization} \label{sec:WSM}
Because of the non-convex objective and constraints,   $\mathbf{P1}$ is not amenable to conventional convex optimization methods. And the joint optimization of $\mathbf{w}$ and $\boldsymbol{\rho}$ is complex.  Hence, we seek to solve  $\mathbf{P1}$ by decoupling the optimization variables, $\mathbf{w}$ and $\boldsymbol{\rho}$, to get two sub-problems. Thus, when  power allocation $\boldsymbol{\rho}$ is fixed, $\mathbf{P1}$  reduces to the following  transmit beamforming optimization problem:
\begin{subequations}\label{P1w_prob}
    \begin{eqnarray}
   \!\!\!\!\!     \mathbf{P1}_{\rm{w}}:
        \underset {\mathbf{w}}{\rm{maximize}} && \sum\nolimits_{k\in \mathcal{K}_{0}} a_k {\rm{log}}_2(1+\gamma_{k}), \label{P1w_obj} \\
        \text{subject to} && \gamma_k \ge \gamma_k^{\rm{th}},  \quad \text{for} \quad k \in \mathcal{K}_{0}, \label{P1w_rateUi}\\ 
        && \eta_b (1-\alpha)  \vert \mathbf{f}^{\rm{H}} \mathbf{w} \vert ^2 p_t \ge p_b,  \label{P1w_BD_EH1}\\
        &&\begin{cases}
      \Vert\mathbf{w}\Vert^2=1, &  \text{Digital,} \\ 
        \vert[\mathbf{w}]_m \vert = \frac{1}{\sqrt{M}}, m \in \mathcal{M} &\text{Analog}, \\ 
 \end{cases} \quad \label{P1w_preco} \\
        && \vert \mathbf{h}_1^{\rm{H}} \mathbf{w} \vert^2 \ge  \ldots \ge \vert \mathbf{h}_K^{\rm{H}} \mathbf{w} \vert^2, \quad \quad
    \end{eqnarray}
\end{subequations}
where $\gamma_k^{\rm{th}}= 2^{R_k^{\rm{th}}} - 1$ and the rate requirement constraint in \eqref{P1_rateUi} is equivalently converted to SINR constraint in \eqref{P1w_rateUi}, since $R_k$ is a non-decreasing function of its argument. Similarly, for a given  $\mathbf{w}$, $\mathbf{P1}$ becomes a power allocation problem at the BS. We formulate the corresponding optimization problem for the power allocation factor as
\begin{subequations}\label{P1r_prob}
    \begin{eqnarray}
        \mathbf{P1_{{\rho}}}:
        \underset {\boldsymbol{\rho}}{\rm{maximize}} && \sum\nolimits_{k\in \mathcal{K}_0} a_k {\rm{log}}_2(1+\gamma_{k}), \label{P1r_obj}\\
        \text{subject to} && \gamma_k \ge \gamma_k^{\rm{th}}, \quad \text{for} \quad k \in \mathcal{K}_0, \label{P1r_rateUi}\\ 
        && \rho_{1} \leq \cdots \leq \rho_{K}, \!\!\quad \text{and} \!\!\quad \sum\nolimits_{k\in \mathcal{K}}{\rho_{k}} = 1. \qquad \label{P1r_power} 
    \end{eqnarray}
\end{subequations}

We note that $\mathbf{P1}_{\rm{w}}$ and $\mathbf{P1_{{\rho}}}$ are not convex because of the non-convexity of the corresponding objective functions and the constraints. We employ the AO technique in which $\mathbf{P1}_{\rm w}$ and $\mathbf{P1_{{\rho}}}$ are alternately maximized until the WSR objective converges. 

The AO technique is an iterative approach for optimizing a function, $f(\mathbf{x})$, jointly over all variables, $\mathbf{x}=\{x_1,\ldots,x_m\}$, by alternating constrained optimizations over the individual subsets of variables \cite{Bezdek2003}. It is also referred to as block relaxation, nonlinear block  Gauss-Seidel,  or block-coordinated descent. This method first separates variables into non-overlapping blocks of one or more variables.  After that,  $f(\mathbf{x})$ is optimized over all variables by alternating through the block variables. Moreover, while optimizing over one block variable, all other blocks are kept fixed at their current values \cite{Bezdek2003}.

\subsubsection{Transmit Beamforming}
 We introduce  $\beta_k$ to replace the SINR terms in  \eqref{P1w_obj} such that $\beta_k\leq \gamma_k$, $\mathbf{P1}_{\rm w}$ can be reformulated to a standard FP problem as 
\begin{subequations}\label{P1w1_prob}
    \begin{eqnarray}
       \mathbf{P1}_{\rm{w}1}:
        \underset {\mathbf{w}, \boldsymbol{\beta}}{\rm{maximize}} && \sum\nolimits_{k \in \mathcal{K}_0} a_k {\rm{log}}_2(1+\beta_{k}), \label{P1w1_obj} \\
        \text{subject to} && \gamma_k^{\rm{th}} \leq \beta_k \leq \frac{A_k(\mathbf{w})}{B_k(\mathbf{w})},  \quad \text{for} \quad k \in \mathcal{K}_0, \qquad  \label{P1w1_rateUi}\\ 
        && \eta_b (1-\alpha)  \vert \mathbf{f}^{\rm{H}} \mathbf{w} \vert ^2 p_t \ge p_b,  \label{P1w1_BD_EH}\\
        &&\begin{cases}
      \Vert\mathbf{w}\Vert^2=1, &  \text{Digital,} \\ 
        \vert[\mathbf{w}]_m \vert = \frac{1}{\sqrt{M}}, m \in \mathcal{M} &\text{Analog}, \\ 
 \end{cases} \label{P1w1_preco}\quad \quad \\
        && \vert \mathbf{h}_1^{\rm{H}} \mathbf{w} \vert^2 \ge  \ldots \ge \vert \mathbf{h}_K^{\rm{H}} \mathbf{w} \vert^2, \qquad \label{P1w1_order}
    \end{eqnarray}
\end{subequations}
where $\boldsymbol{\beta}=[\beta_0,\ldots, \beta_K]^{\rm{T}}$. Here, $A_k(x)$ and $B_k(x)$ are the numerator and denominator of the $k$-th SINR term as functions  $x=\mathbf{w}$. We note that, per  \eqref{Eqn:rate_u2_min},  to meet the minimum condition of $\gamma_k,  k \in \mathcal{K}$, we ensure that both $\gamma_{k}^{k}$ and $\gamma_{i}^{k} (i \in \mathcal{K}_k)$ satisfy the SINR threshold requirement. Next, $\mathbf{P1}_{\rm{w}1}$ can be seen as a two-part optimization problem: (i) an outer optimization over $\mathbf{w}$ with fixed $\boldsymbol{\beta}$ and (ii) an inner optimization over $\boldsymbol{\beta}$ with fixed $\mathbf{w}$. We give the inner optimization problem as
\begin{subequations}\label{P1w2_prob}
    \begin{eqnarray}
        \mathbf{P1}_{\rm{w}2}:
        \underset { \boldsymbol{\beta}}{\rm{maximize}} && \sum\nolimits_{k\in \mathcal{K}_0} a_k {\rm{log}}_2(1+\beta_{k}), \label{P1w2_obj} \\
        \text{subject to} && \gamma_k^{\rm{th}} \leq \beta_k \leq \frac{A_k(\mathbf{w})}{B_k(\mathbf{w})},  \quad \text{for} \quad k \in \mathcal{K}_0.\qquad  \label{P1w2_rateUi}
    \end{eqnarray}
\end{subequations}

The inner optimization problem in \eqref{P1w2_prob} is convex in $\boldsymbol{\beta}$ and holds strong duality \cite{Shen2018}. {Hence, the trivial solution to it is that $\beta_k$ satisfies \eqref{P1w2_rateUi} with equality, i.e., $\beta_k^{o}=A_k(\mathbf{w})/B_k(\mathbf{w})$.}   To tackle the logarithm in the objective function of $\mathbf{P1}_{\rm{w}2}$, we apply the Lagrangian dual transform \cite{Shen2018}, and the corresponding Lagrangian function is given as
\begin{eqnarray} \label{LD_func_w}
    L(\boldsymbol{\beta},\boldsymbol{\lambda}) &=& \sum\nolimits_{k\in \mathcal{K}_0}  a_k {\rm{log}}_2(1+\beta_{k}) \nonumber \\
    &-&\sum\nolimits_{k\in \mathcal{K}_0} \lambda_k \left( \beta_k -\frac{A_k(\mathbf{w})}{B_k(\mathbf{w})} \right),\quad
\end{eqnarray}
where $\boldsymbol{\lambda}=[\lambda_0, \ldots,\lambda_K]^{\rm{T}}$ is the dual variable vector introduced for each inequality constraint in \eqref{P1w2_rateUi}. From the strong duality, we can equivalently reformulate $\mathbf{P1}_{\rm{w}2}$ to a dual problem as
\begin{eqnarray}\label{P1w3_prob}
    \mathbf{P1}_{\rm{w}3}:
    \underset {\boldsymbol{\lambda} \succeq 0}{\rm{minimize}} \quad \underset {\boldsymbol{\beta} }{\rm{maximize}} \quad L(\boldsymbol{\beta},\boldsymbol{\lambda}).
\end{eqnarray}
By evaluating the first-order condition $\partial L(\boldsymbol{\beta},\boldsymbol{\lambda})/ \partial \beta_k, k \in \mathcal{K}_0$ and using the trivial solution to the inner optimization problem, we obtain the optimal solution of $\lambda_k$ as
\begin{eqnarray}\label{lambda_opt}
    \lambda_k^{o} = \frac{a_k B_k(\mathbf{w})}{A_k(\mathbf{w})+B_k(\mathbf{w})},  k \in \mathcal{K}_0.
\end{eqnarray}
We note that $\lambda_k \geq 0$ is automatically satisfied in this case. From \eqref{P1w2_prob} and \eqref{lambda_opt}, we can reformulate the inner optimization as
\begin{eqnarray}\label{P1w4_prob}
	\mathbf{P1}_{\rm{w}4}:
  \underset {\boldsymbol{\beta} }{\rm{maximize}} \quad L(\boldsymbol{\beta},\boldsymbol{\lambda}^o).
\end{eqnarray}
Furthermore, we can prove that when coupled with the outer optimization over $\mathbf{w}$ and after several mathematical interpretations, the solution to $\mathbf{P1}_{\rm{w}4}$ also satisfies $\mathbf{P1}_{\rm{w}1}$ \cite{Shen2018}. Additionally, the beamforming vector, $\mathbf{w}$, is obtained by solving the feasibility problem over $\mathbf{w}$ for fixed $\boldsymbol{\beta}$ \eqref{P1w1_prob} (Appendix \ref{apx:Convexify}). The detailed steps of the proposed solutions are outlined in Algorithm \ref{Algo1}.

\begin{rem}\label{rem:CM}
When the constant-modulus beamforming is used in $\mathbf{P1}_{\rm{w}1}$, the constraint \eqref{P1w1_preco} is equivalent to the phase-shift constraint in hybrid precoding problems \cite{Park2017}. Hence, similar to \cite{Park2017}, this can be handled by first relaxing the constant-modules condition and then solving for the optimal precoder. Thereby, once the precoder is obtained, the constant-modules condition can be imposed such that $[\mathbf{w}^o]_m= \frac{1}{\sqrt{M}} \Exp{j \angle([\mathbf{w}_{rx}^o]_m)}, m \in \mathcal{M}$, where $\mathbf{w}_{rx}^o$ is the optimal precoder with relaxed the constant-modules condition \cite{Park2017}.
\end{rem}

\linespread{1.0}
\begin{algorithm}[hbt!]
\caption{: WSRMax for transmit beamforming.}\label{Algo1}
{\small{\begin{algorithmic}
    \renewcommand{\algorithmicrequire}{\textbf{Initialization:}}
    \renewcommand{\algorithmicensure}{\textbf{Repeat}}
    \Require Initialize $\mathbf{w}$ to a feasible value.
    \Ensure
    \State \textbf{Step 1}: Update $\boldsymbol{\lambda}$ by \eqref{lambda_opt}.
    \State \textbf{Step 2}: Update $\boldsymbol{\beta}$ by solving $\mathbf{P1}_{\rm{w}4}$ in \eqref{P1w4_prob}.
    \State \textbf{Step 3}: Update $\mathbf{w}$ by solving feasibility problem  over $\mathbf{w}$ for fixed  $\boldsymbol{\beta}$ \eqref{P1w1_prob}.
\end{algorithmic}
\textbf{Until} the value of the objective function converges.\\
\textbf{Output:} The optimal beamforming vector $\mathbf{w}^o$.}}
\end{algorithm}

\subsubsection{Power Allocation}
By following a similar approach to $\mathbf{P1}_{\rm{w}}$,  we introduce a new variable $\theta_k$ for replacing each SINR term in the objective function \eqref{P1r_obj}, and $\mathbf{P1_{{\rho}}}$ is reformulated as
\begin{subequations}\label{P1r1_prob}
    \begin{eqnarray}
   \!\!\!\!\!     \mathbf{P1}_{{\rho} 1}:
        \underset {\boldsymbol{\rho},\boldsymbol{\theta}}{\rm{maximize}} && \sum\nolimits_{k\in \mathcal{K}_0} a_k {\rm{log}}_2(1+\theta_{k}), \label{P1r1_obj}\\
        \text{subject to} && \gamma_k^{\rm{th}} \leq \theta_k \leq \frac{A_k(\boldsymbol{\rho})}{B_k(\boldsymbol{\rho})} , \quad \text{for} \quad k \in \mathcal{K}_0, \label{P1r1_rateUi}\quad\\ 
        &&  \!\rho_{1} \leq \cdots \leq \rho_{K}, \!\!\quad \text{and} \!\!\quad \sum\nolimits_{k\in \mathcal{K}}\!{\rho_{k}} = 1\!,\label{P1r1_power} \quad
    \end{eqnarray}
\end{subequations}
where $\boldsymbol{\theta}=[\theta_0,\ldots,\theta_K]^{\rm{T}}$. Similar to  $ \mathbf{P1}_{\rm{w}1}$, $ \mathbf{P1}_{{\rho} 1}$ is also a two part optimization problem. For a fixed $\boldsymbol{\rho}$, the inner optimization problem is given as
\begin{subequations}\label{P1r2_prob}
\begin{eqnarray}
    \mathbf{P1}_{{\rho} 2}:
    \underset {\boldsymbol{\theta}}{\rm{maximize}} && \sum\nolimits_{k\in \mathcal{K}_0} a_k {\rm{log}}_2(1+\theta_{k}), \label{P1r2_obj}\\
    \text{subject to} && \gamma_k^{\rm{th}} \leq \theta_k \leq \frac{A_k(\boldsymbol{\rho})}{B_k(\boldsymbol{\rho})} , \quad \text{for} \quad k \in \mathcal{K}_0.\qquad  \label{P1r2_rateUi}
\end{eqnarray}
\end{subequations}
{Here, the trivial solution to $\theta_k$ satisfies \eqref{P1r2_rateUi} with equality, i.e., $\theta_k^{o}=A_k(\boldsymbol{\rho})/B_k(\boldsymbol{\rho})$.}  By introducing a dual variable, $y_k$, the  corresponding Lagrangian function is given as

\begin{eqnarray} \label{LD_func_r}
     L(\boldsymbol{\theta},\mathbf{y}) \!=\! \sum\limits_{k\in \mathcal{K}_0} \!a_k {\rm{log}}_2(1+\theta_{k}) -\!\sum\limits_{k\in \mathcal{K}_0} \!y_k \left(\theta_k -\frac{A_k(\boldsymbol{\rho})}{B_k(\boldsymbol{\rho})} \right)\!,\quad
\end{eqnarray}
where $\mathbf{y}=[y_0,\ldots,y_K]^{\rm{T}}$. Due to strong duality, $\mathbf{P1}_{{\rho} 2}$ can be equivalently reformulated as
\begin{eqnarray}\label{P1r3_prob}
    \mathbf{P1}_{{\rho} 3}:
    \underset {\mathbf{y} \succeq 0}{\rm{minimize}} \quad \underset {\boldsymbol{\beta} }{\rm{maximize}} \quad L(\boldsymbol{\theta},\mathbf{y}).
\end{eqnarray}
Next, we evaluate the first-order condition $\partial L(\boldsymbol{\theta},\mathbf{y})/ \partial \theta_k, k \in \mathcal{K}_0$ to obtain the optimal value of the dual variable $y_k$ as
\begin{eqnarray}\label{y_opt}
    y_k^{o} = \frac{a_k B_k(\boldsymbol{\rho})}{A_k(\boldsymbol{\rho})+B_k(\boldsymbol{\rho})}, k \in \mathcal{K}_0.
\end{eqnarray}
Hence, $\mathbf{P1}_{{\rho} 2}$ is equivalent to
\begin{eqnarray}\label{P1r4_prob}
    \mathbf{P1}_{{\rho} 4}:
    \underset {\boldsymbol{\theta} }{\rm{maximize}} \quad L(\boldsymbol{\theta},\mathbf{y}^o).
\end{eqnarray}
The proposed algorithm for solving the power allocation problem is given in Algorithm \ref{Algo2}.

\linespread{1.0}
\begin{algorithm}[hbt!]
\caption{: WSRMax for power allocation.}\label{Algo2}
 {\small{\begin{algorithmic}
        \renewcommand{\algorithmicrequire}{\textbf{Initialization:}}
        \renewcommand{\algorithmicensure}{\textbf{Repeat}}
        \Require Initialize $\boldsymbol{\rho}$ to a feasible value.
        \Ensure
        \State \textbf{Step 1}: Update $\mathbf{y}$ by \eqref{y_opt}.
        \State \textbf{Step 2}: Update $\boldsymbol{\theta}$ by solving $\mathbf{P1}_{{\rho} 4}$ in \eqref{P1r4_prob}.
        \State \textbf{Step 3}: Update $\boldsymbol{\rho}$ by solving feasibility problem over $\boldsymbol{\rho}$ for fixed  $\boldsymbol{\theta}$ \eqref{P1r1_prob}.
    \end{algorithmic}
\textbf{Until} the value of the objective function converges.\\
\textbf{Output:} The optimal power allocation factor $\boldsymbol{\rho}^o$. }}
\end{algorithm}

\begin{rem}\label{rem:overal_algo_WSR}
Algorithms \ref{Algo1} and \ref{Algo2} outline the proposed optimization approaches for solving $\mathbf{w}$ by fixing $\boldsymbol{\rho}$ and for solving $\boldsymbol{\rho}$ by fixing $\mathbf{w}$ after the original problem, $\mathbf{P1}_{\rm{w}}$, is separated into two sub-problems. We begin by quantifying the SINR of $U_k, k \in \mathcal{K}$, and the tag as we initialize $\mathbf{w}$ and $\boldsymbol{\rho}$ to feasible values, and then better solutions for $\mathbf{w}$ and $\boldsymbol{\rho}$ are updated in each iteration. The procedure is repeated until there is no further improvement. At this point, it is terminated by a condition such that the increment of the normalized objective function is smaller than $\epsilon=10^{-3}$.
\end{rem}

\subsection{Transmit Power Minimization}\label{Sec:transmit_min}
Since $R_k, k \in \mathcal{K}_0$  in  $\mathbf{P2}$ is a non-decreasing function of its argument $\gamma_k$, the rate requirements in \eqref{P2_rateUi} of $\mathbf{P2}$ can be replaced with the corresponding SINR constraints. Thereby, the TPMin problem in $\mathbf{P2}$ is reformulated as

\begin{subequations}\label{P21_prob}
    \begin{eqnarray}
        \mathbf{P2}_{1}:
        \underset {\mathbf{w},p_t,\boldsymbol{\rho}}{\rm{minimize}} && p_t, \\
        \text{subject to} && 
        \gamma_k \ge \gamma_k^{\rm{th}},   \quad \text{for} \quad k \in \mathcal{K}_0,\label{P21_rateUi}\\ 
        &&\eta_b (1-\alpha)  \vert \mathbf{f}^{\rm{H}} \mathbf{w} \vert ^2 p_t \ge p_b,  \label{P21_BD_EH}\\
        &&  \rho_{1} \leq \cdots \leq \rho_{K}, \!\!\quad \text{and} \!\!\quad \sum\nolimits_{k\in \mathcal{K}}{\rho_{k}} = 1,\label{P21_power} \qquad \\
        &&\Vert \mathbf{w}\Vert^2=1,  \label{P21_preco} \\
        && \vert \mathbf{h}_1^{\rm{H}} \mathbf{w} \vert^2 \ge  \ldots \ge \vert \mathbf{h}_K^{\rm{H}} \mathbf{w} \vert^2.\quad 
    \end{eqnarray}
\end{subequations}

Because of the non-convexity in the constraint functions of $\mathbf{P2}_1$, similar to Section \ref{sec:WSM}, we aim to obtain a sub-optimal solution for $\mathbf{P2}_1$ by decoupling the optimization variables. Thus, for a given power allocation factor $\boldsymbol{\rho}$, $\mathbf{P2}_1$ becomes a transmit power/precoder minimization problem as 
\begin{subequations}\label{P2p_prob}
    \begin{eqnarray}
        \mathbf{P2}_{\rm{p}}:
        \underset {\mathbf{w},p_t}{\rm{minimize}} && p_t, \\
        \text{subject to} && 
        \gamma_k \ge \gamma_k^{\rm{th}},   \quad \text{for} \quad k \in \mathcal{K}_0,\label{P2p_rateUi}\\ 
        &&\eta_b (1-\alpha)  \vert \mathbf{f}^{\rm{H}} \mathbf{w} \vert ^2 p_t \ge p_b,  \label{P2p_BD_EH}\\
        &&\Vert \mathbf{w}\Vert^2=1,  \label{P2p_preco} \\
        &&\vert \mathbf{h}_1^{\rm{H}} \mathbf{w} \vert^2 \ge  \ldots \ge \vert \mathbf{h}_K^{\rm{H}} \mathbf{w} \vert^2. \qquad
    \end{eqnarray}
\end{subequations}
Next, for given a $p_t$ and a  $\mathbf{w}$ $\mathbf{P2}_1$ is reduced to a feasibility problem as follows:
\begin{subequations}\label{P2r_prob}
    \begin{eqnarray}
        \mathbf{P2}_{\rho}:
        \rm{find} && \boldsymbol{\rho}, \\
        \text{subject to} && 
        \gamma_k \ge \gamma_k^{\rm{th}},   \quad \text{for} \quad k \in \mathcal{K}_0,\label{P2r_rateUi}\\ 
        &&  \rho_{1} \leq \cdots \leq \rho_{K}, \quad \text{and} \quad \sum\nolimits_{k\in \mathcal{K}}{\rho_{k}} = 1.\quad\label{P2r_power} 
    \end{eqnarray}
\end{subequations}

\subsubsection{Transmit Power/Precoder Minimization}
By defining $\mathbf{v}\triangleq \sqrt{p_t} \mathbf{w}$,  $\mathbf{W}\triangleq \mathbf{v} \mathbf{v}^{\rm{H}}$, and $\mathbf{A}\triangleq \mathbf{a} \mathbf{a}^{\rm{H}}$, where $\mathbf{a} \!\in\! \{ \mathbf{h}_{k}, \mathbf{g}_{k} \}, k \!\in \! \mathcal{K}$, we can rewrite the SINR terms in \eqref{Eqn:rate_u2_min} and \eqref{Eqn:upper} respectively as \eqref{SDR_rate},
\begin{figure*}
\begin{eqnarray}\label{SDR_rate}
    \gamma_{i}^{k } =  \frac{\rho_k \Tr(\mathbf{H}_i \mathbf{W}) }{ \Tr (\mathbf{H}_i \mathbf{W})\sum\nolimits_{j \in \mathcal{K}_k} {\rho_j } + \Tr (\mathbf{H}_i \mathbf{W}) \sum\nolimits_{j \in \mathcal{K}_k'} {\rho_j \left( 2-2 \xi_j \right)} +\alpha \Tr ( \mathbf{G}_i \mathbf{W})  + \sigma^2}
\end{eqnarray}
  \hrulefill
  \vspace{-3mm}
\end{figure*}
and
\begin{eqnarray}
 \gamma_{0} &=& \frac{\alpha \Tr(\mathbf{G}_1 \mathbf{W})}{2\left(\Tr (\mathbf{H}_1 \mathbf{W}) \sum_{j \in \mathcal{K}} {\rho_j \left( 2-2 \xi_j \right)}+\sigma^2\right)}.
\end{eqnarray}
Additionally, $\gamma_k = {\rm{min}}(\gamma_{k}^{k},\gamma_{i}^{k})$.  Then, $\mathbf{P2}_{\rm{p}}$ is reformulated into the following equivalent problem:
\begin{subequations}\label{P2p1_prob}
    \begin{eqnarray}
        \mathbf{P2}_{\rm{p}1}:
        \underset {\mathbf{W}}{\rm{minimize}} && \Tr(\mathbf{W}), \\
        \text{subject to} && 
        \gamma_k \ge \gamma_k^{\rm{th}},   \quad \text{for} \quad k \in \mathcal{K}_0,\label{P2p1_rateUi}\\ 
        &&\eta_b (1-\alpha)  \Tr ( \mathbf{F} \mathbf{W})  \ge p_b,  \label{P2p1_BD_EH}\\
        && \Tr(\mathbf{W}) \leq p_{\rm{max}}, \label{P2p1_max_power}\\
        && \rm{Rank}(\mathbf{W}) =1,  \label{P2p1_rank} \\
        && \Tr(\mathbf{H}_1 \mathbf{W}) \ge \ldots \ge \Tr (\mathbf{H}_K \mathbf{W} ),
    \end{eqnarray}
\end{subequations}
where $p_{\rm{max}}$ is the maximum allowable transmit power at the BS. By relaxing the non-convex rank-one constraint in \eqref{P2p1_rank}, $\mathbf{P2}_{\rm{p}1}$ can be reformulated to the following  SDR problem \cite{BoydConvex2004,Long2020}:
\begin{subequations}\label{P2p2_prob}
    \begin{eqnarray}
        \mathbf{P2}_{\rm{p}2}:
        \underset {\mathbf{W}}{\rm{minimize}} && \Tr(\mathbf{W}), \\
        \text{subject to} && 
        \gamma_k \ge \gamma_k^{\rm{th}},   \quad \text{for} \quad k \in \mathcal{K}_0,\label{P2p2_rateUi}\\ 
        &&\eta_b (1-\alpha)  \Tr ( \mathbf{F} \mathbf{W})  \ge p_b,  \label{P2p2_BD_EH}\\
        && \Tr(\mathbf{W}) \leq p_{\rm{max}}, \label{P2p2_max_power}\\
        && \Tr(\mathbf{H}_1 \mathbf{W}) \ge \ldots \ge \Tr (\mathbf{H}_K \mathbf{W}).
    \end{eqnarray}
\end{subequations}
This SDR problem, $\mathbf{P2}_{\rm{p}2}$, is a convex optimization problem and can be solved by using semi-definite programming (SDP) via CVX Matlab \cite{BoydConvex2004,Michaelcvx}. Then, if the solution to this SDR problem is of rank one, i.e., $\mathbf{W}^o=\mathbf{v}^o (\mathbf{v}^o)^{\rm{H}}$, the solutions to  $\mathbf{P2}_{\rm{p}1}$ is $p_t^o=\Tr(\mathbf{S})$ and $\mathbf{w}^o=\mathbf{v}^o/\sqrt{p_t^o}$, where $\mathbf{S}$ is obtained by computing the singular value decomposition (SVD) of $\mathbf{W}^o$ as $\mathbf{W}^o=\mathbf{U} \mathbf{S} \mathbf{U}^{\rm{H}}$. Otherwise, we employ the Gaussian randomization-based technique to achieve an approximate sub-optimal solution to $\mathbf{P2}_{\rm{p}1}$ \cite{Sidiropoulos2006}. The steps to find the solution to $\mathbf{P2}_{\rm{p}1}$ are summarized in Algorithm \ref{Algo3}.

\linespread{1.0}
\begin{algorithm}[hbt!]
\caption{: TPMin  for power/precoder minimization.}\label{Algo3}
{\small{       \begin{algorithmic}
        \renewcommand{\algorithmicrequire}{\textbf{Initialization:}}
        \Require  Begin - CVX.
        \State \textbf{Step 1}: Solve the convex  problem $\mathbf{P2}_{\rm{p}2}$ in \eqref{P2p2_prob}.
        \State \textbf{Step 2}: SVD  $\mathbf{W}^o$ as $\mathbf{W}^o=\mathbf{U} \mathbf{S} \mathbf{U}^{\rm{H}}$, where $\mathbf{U}=[\mathbf{u}_1, \ldots, \mathbf{u}_M]$. 
        \State End - CVX.
        \If{$\rm{Rank}(\mathbf{W}^o)=1$,} 
        \State \textbf{return}: $p_t^o=\Tr(\mathbf{S})$ and $\mathbf{w}^o=\mathbf{u}_1$.
        \Else
        \For{$d=1,\ldots,D$}
            \State 1: Generate random  $\mathbf{v}_d = \mathbf{U} \mathbf{S}^{1/2} \mathbf{e}_d$, where $\mathbf{e}_d = [\Exp{j\phi_1}, \ldots, \Exp{j\phi_M}]^{\rm{H}}$ and $\phi_i \sim \mathcal{U}(0,2\pi)$.
            \State 2: Check if {$\mathbf{P2}_{\rm{p}2}$ is feasible with $\mathbf{v}_d$.} 
        \EndFor
        \State \textbf{return}: $p_t^o=\Vert \mathbf{v}^o \Vert^2$ and $\mathbf{w}^o=\mathbf{v}^o/\sqrt{p_t^o}$, where  $\mathbf{v}^o= \underset {d=1,\ldots,D}{\rm{arg \,min}} \,\, \mathbf{v}_d$.
        \EndIf 
    \end{algorithmic}
\textbf{Output:} Optimal transmit power $p_t^o$ and  precoder $\mathbf{w}^o$. }}
\end{algorithm}

\subsubsection{Transmit Power Allocation}
The power allocation problem, $\mathbf{P2}_{\rho}$, resembles the relay beamforming optimization problem for the multi-antenna relay broadcast channel \cite{Zhang2009}. Hence,  we transform it into a second-order cone programming (SOCP) optimization problem and solve it via CVX Matlab \cite{Zhang2009,BoydConvex2004,Michaelcvx}.

\begin{rem}\label{rem:overal_algo_PM}
We also split  $\mathbf{P2}_{1}$ into two sub-problems: (i) solve for $p_t$ and $\mathbf{w}$ by fixing $\boldsymbol{\rho}$ (i.e., Algorithm \ref{Algo3}) and (ii) solve for $\boldsymbol{\rho}$ by fixing $p_t$ and $\mathbf{w}$ (i.e., $\mathbf{P2}_{\rho}$). We first establish $p_t$, $\mathbf{w}$, and $\boldsymbol{\rho}$ to feasible values, and then update better solutions for them in each iteration. The process continues until there is no further improvement. Specifically, it terminates when the increment of the normalized objective function is less than $\epsilon=10^{-3}$.
\end{rem}

\subsection{Computational Complexity}

The proposed algorithms have multiple stages. In WSRMax  (i.e., Algorithm \ref{Algo1} and Algorithm \ref{Algo2}), the outer loop comprises two sub-problems for optimizing $\mathbf{w}$ and $\boldsymbol{\rho}$.   The main complexity of these two lies in step 3. As CVX Matlab applies an SDPT3 solver to handle these optimization problems, the computational complexities are $\mathcal{O}((K+1)^3M^3)$ and $\mathcal{O}((K+1)^3)$, respectively \cite{Ben2001book}. 
Hence, the proposed solution for WSRMax has a total complexity of $\mathcal{O}(I_o^W(I_w (K+1)^3 M^3+  I_{\rho} (K+1)^3 ))$, where $I_w$, $I_{\rho}$, and $I_o^W$ are the iteration numbers of Algorithm \ref{Algo1}, Algorithm \ref{Algo2}, and the overall algorithm (outer loop), respectively.

The  problem in $\mathbf{P2}_{\rm{p}2}$ \eqref{P2p2_prob} is conventional SDP, solvable via  interior point methods \cite{Karipidis2007}. 
In Algorithm \ref{Algo3}, for  a  $M\times M$ matrix $\mathbf{W},$  those methods  require $\mathcal{O}(\sqrt{M} \log(1/\kappa))$ iterations, with each iteration taking $\mathcal{O}(M^6)$ arithmetic operations in the worst case, where $\kappa$ is the interior-point algorithm's precision \cite{Karipidis2007,Long2020}.

\section{Simulation Results}\label{sim}
Herein, we provide simulation examples to evaluate the benefits/harms of symbiosis (Fig. 2).  We adopt the 3GPP UMi model to model the large-scale fading $\zeta_{a}$ \eqref{pathloss} with $f_c =  \qty{3}{\GHz}$ operating frequency \cite[Table B.1.2.1]{3GPP2010}. 
The  AWGN variance, $\sigma^2$, is modeled as $\sigma^2=10\log_{10}(N_0 B N_f)$ dBm, where $N_0=\qty{-174}{\dB m/\Hz}$, $B =  \qty{10}{\MHz}$ is the bandwidth, and $N_f = \qty{10}{\dB}$ is the noise figure. Moreover, $M=32$,  $a_k = 1/(K+1), k \in \mathcal{K}_0$, $p_b = \qty{-20}{\dB m}$ and $\eta_b = 0.6$. Furthermore, we consider the fixed quality of estimation (or constant correlation coefficient), i.e., $\xi_j=\xi, \forall j$.

To comparatively evaluate the network performance (Fig. 1),  we also consider a traditional  (i.e., non-symbiotic)   NOMA network without the tag.  For that,   the WSRMax and TPMin problems are solved as in  Section \ref{sec:WSM} and Section \ref{Sec:transmit_min}, respectively.

\noindent\textbf{Weighted Sum-Rate Maximization:}
Herein,  we consider two primary NOMA users ($K=2$) and investigate their rate performances and the tag rate, and the EH constraint at the tag through the proposed solutions for the  WSRMax problem (Section \ref{sec:WSM}), under perfect and imperfect SIC. {We consider the tag's target rates as $R_0^{\rm{th}}=\log_2(1+p_t/100)$,  $R_1^{\rm{th}}=\log_2(1+p_t)$, and $R_2^{\rm{th}}=\log_2(1+p_t/10)$. }
If constant rate thresholds are set, in low transmit powers, the program will not converge. Therefore, the rate thresholds must be functions of the transmit power. At different power levels, the devices achieve different rates, which are supported by beamforming \cite{Kudathanthirige2019,Perera2022}.

We consider a  BS with   $M=32$ antennas and digital/analog beamforming capacity  (Section \ref{Sec:system_model} and Remark \ref{rem:CM}). As benchmarks, we also consider digital weighted maximum ratio transmission (MRT) and random beamforming designs. For the latter, a complex Gaussian random vector is selected to satisfy the NOMA constraint \eqref{NOMA_c}. Hence, it does not require CSI, and the beam can be focused in any direction, resulting in poor performance \cite{Jaehak2003}. Weighted MRT is designed as $0.5 \mathbf{h}_1/\Vert \mathbf{h}_1 \Vert + 0.5 \mathbf{h}_2/\Vert \mathbf{h}_2 \Vert $. It directs the beam toward the primary users and thus necessitates accurate CSI \cite{Jiawei2018}. Moreover, for both designs, we consider $\rho_1 = \num{0.3}$ to satisfy \eqref{P1_power}. We assume $d_f=\qty{5}{\m}$, $d_{h_1}= d_{h_2}=\qty{12}{\m}$, $d_{q_1}=\qty{8}{\m}$, $d_{q_2}=\qty{10}{\m}$, and $\alpha=0.6$. { These parameter values are taken from \cite{Xu2021NOMA, Zhuang2022, Galappaththige2023}}.

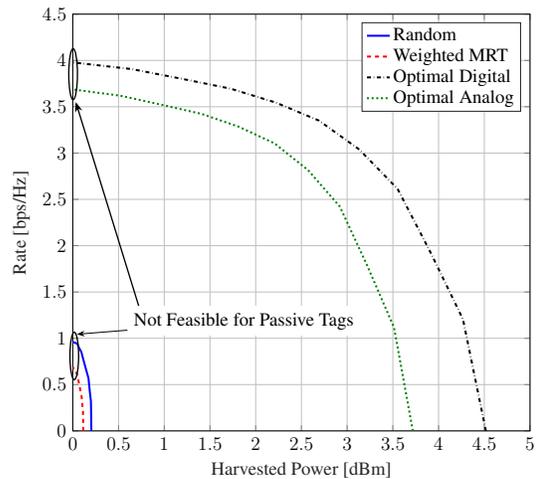
\begin{figure}[!t]\centering \vspace{-4mm}
    \fontsize{14}{14}\selectfont 
    \resizebox{.58\totalheight}{!}{
%
%
\definecolor{mycolor1}{rgb}{0.00000,0.49804,0.00000}%
\begin{tikzpicture}

\begin{axis}[%
width=4.755in,
height=4.338in,
at={(0.798in,0.586in)},
scale only axis,
xmin=0,
xmax=5,
xlabel style={font=\color{white!15!black}},
xlabel={Harvested Power [dBm]},
ymin=0,
ymax=4.5,
ylabel style={font=\color{white!15!black}},
ylabel={Rate [bps/Hz]},
axis background/.style={fill=white},
xmajorgrids,
ymajorgrids,
legend style={legend cell align=left, align=left, draw=white!15!black}
]
\addplot [color=blue, line width=1.5pt]
  table[row sep=crcr]{%
0.202793402555678	0\\
0.198072316056824	0.32116840968435\\
0.18553607411175	0.425626431321559\\
0.171198218906408	0.56864358056023\\
0.152355073632982	0.636404914207249\\
0.131357813793844	0.711029742334465\\
0.110853321693741	0.786065755915417\\
0.0931344889181061	0.851630661310536\\
0.0669136636123128	0.900707726452898\\
0.0474489326698616	0.942700864780089\\
0	0.960471071905166\\
};
\addlegendentry{Random}

\addplot [color=red, dashed, line width=1.5pt]
  table[row sep=crcr]{%
0.114408771277858	0\\
0.111405485085662	0.209963190076944\\
0.103876477822265	0.280778623551728\\
0.0954447782189058	0.37799975167206\\
0.084559451923951	0.428217312357179\\
0.0728242194428599	0.485242766390282\\
0.0615047987978088	0.543336218435036\\
0.0518359650172725	0.595028700726414\\
0.0373859103650603	0.63423965736147\\
0.0267415539088011	0.668364547661803\\
0	0.682160423776945\\
};
\addlegendentry{Weighted MRT}

\addplot [color=black, dashdotted, line width=1.5pt]
  table[row sep=crcr]{%
4.51752233306906	0\\
4.26693750595103	1.20443011806174\\
3.91251123125467	1.92628575205314\\
3.55090929081345	2.61387984578672\\
3.13250430622777	3.03868691798865\\
2.69127627241967	3.35070113491731\\
2.21568564559909	3.54654448628676\\
1.72475604478117	3.69544739083696\\
1.18285351860858	3.80359387343243\\
0.643047500075918	3.90685752244131\\
0	3.9787028816452\\
};
\addlegendentry{Optimal Digital}

\addplot [color=mycolor1, dotted, line width=1.5pt]
  table[row sep=crcr]{%
3.71861521970522	0\\
3.51268349458626	1.11640416356561\\
3.21938542976967	1.78553345722224\\
2.91974779359415	2.42270506034042\\
2.57234113964414	2.81619001911722\\
2.20568256468298	3.10483261799002\\
1.81145488223055	3.28577423946948\\
1.40571913236161	3.42330645354991\\
0.961633724596082	3.52356699906046\\
0.519686501160898	3.61932823878582\\
0	3.68716074314364\\
};
\addlegendentry{Optimal Analog}

\end{axis}

\begin{axis}[%
width=6.135in,
height=5.323in,
at={(0in,0in)},
scale only axis,
xmin=0,
xmax=1,
ymin=0,
ymax=1,
axis line style={draw=none},
ticks=none,
axis x line*=bottom,
axis y line*=left
]
\draw [black, line width=1.0pt] (axis cs:0.1305,0.807411) ellipse [x radius=0.00734314, y radius=0.0498628];
\draw [black, line width=1.0pt] (axis cs:0.132461,0.256404) ellipse [x radius=0.00734314, y radius=0.0464318];
\draw[-{Stealth}, color=black, line width=1.0pt] (axis cs:0.249,0.327) -- (axis cs:0.135,0.753);
\draw(axis cs:0.42,0.325) node[fill=white] {Not Feasible for Passive Tags} edge[-Stealth] (axis cs:0.138,0.299);
\end{axis}

\end{tikzpicture}%
}
    \caption{Tag's harvested power and rate trade-off, $p_t=\qty{20}{\dB m}$.}\vspace{-1mm}
    \label{fig:EH}
\end{figure}

\subsection{How Does  Symbiosis  Benefit the Tag?}
The tag must survive with EH. Is it possible to increase the symbiosis benefits of the tag by BS beamforming design? What is the effect of the tag's reflection coefficient on the added benefits?

To shed light on those questions, Fig. \ref{fig:EH} plots the trade-off between the harvested energy and the tag rate as a function of the $\alpha$  with different transmit beamforming vectors, for constant   $p_t =\qty{20}{\dB m}$. As $\alpha$ approaches \num{1}, the tag reflects most of the received power and achieves the maximum rate.   However, this is not a feasible operating point for a passive tag, which must harvest energy. However,  when $\alpha$ approaches \num{0}, the tag harvests most of the received power, and the rate becomes infinitesimal since the tag does not reflect any signal. Hence, for given $\alpha$, the amount of achieved rate and harvested power traverse this trade-off curve. Because the tag must harvest a minimum power, we need   $\alpha \in (0,1)$.

 Fig.~\ref{fig:EH} shows that our BS beamforming design significantly enhances the symbiotic benefits.   Unlike the proposed designs, the conventional random and MRT beamforming does not achieve symbiotic gains for the tag.   Moreover,   analog beamforming can significantly decrease the hardware complexity and power consumption compared to the digital one.  It requires just one RF chain, whereas the digital one requires one RF chain per antenna, \num{32} RF chains.

\begin{figure}[!t]\centering \vspace{-2mm}
    \fontsize{14}{14}\selectfont 
    \resizebox{.56\totalheight}{!}{\input{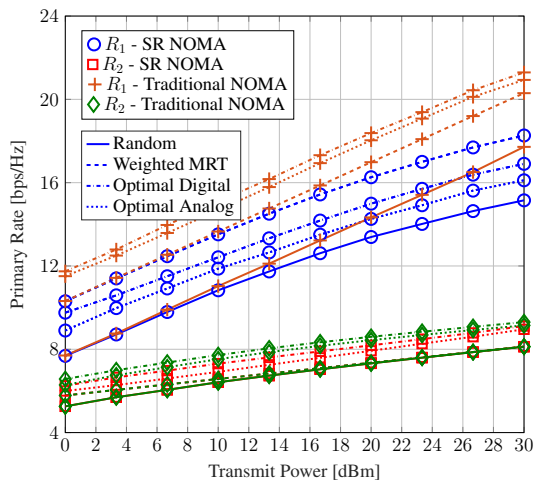}}
    \caption{The  rate of the users versus transmit power.}\vspace{-1mm}
    \label{fig:rate_primary}
\end{figure}
\subsection{Does Symbiosis Have Rate  Benefits?}
 We will investigate this fundamental question. Further, would it be possible to enhance the symbiosis benefits? In this regard, we will find that our algorithm is advantageous and leads to desirable results by maximizing the benefits and minimizing the harms.

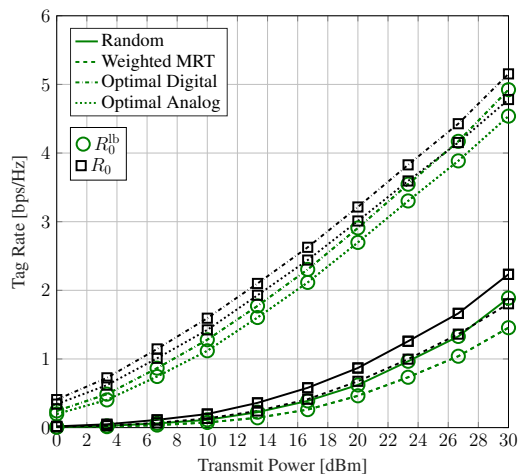
\begin{figure}[!t]\centering \vspace{-2mm}
    \fontsize{14}{14}\selectfont 
    \resizebox{.54\totalheight}{!}{
%
%
\begin{tikzpicture}

\begin{axis}[%
width=4.755in,
height=4.338in,
at={(0.798in,0.586in)},
scale only axis,
xmin=0,
xmax=30,
xlabel style={font=\color{white!15!black}},
xlabel={Transmit Power [dBm]},
ymin=0,
ymax=6,
ylabel style={font=\color{white!15!black}},
ylabel={Tag Rate [bps/Hz]},
axis background/.style={fill=white},
xmajorgrids,
ymajorgrids,
legend style={at={(0.03,0.97)}, anchor=north west, legend cell align=left, align=left, draw=white!15!black}
]
\addplot [color=black!50!green, line width=1.5pt]
  table[row sep=crcr]{%
0	0.00900851780559306\\
3.33333333333333	0.0248175413326591\\
6.66666666666667	0.0615063942714621\\
10	0.112899095460448\\
13.3333333333333	0.224876038494179\\
16.6666666666667	0.38874289687214\\
20	0.62109696162678\\
23.3333333333333	0.966270587480445\\
26.6666666666667	1.32807681821275\\
30	1.88750682644676\\
};
\addlegendentry{Random}

\addplot [color=black!50!green, line width=1.5pt, only marks, mark size=5.0pt, mark=o, mark options={solid, black!50!green}, forget plot]
  table[row sep=crcr]{%
0	0.00900851780559306\\
3.33333333333333	0.0248175413326591\\
6.66666666666667	0.0615063942714621\\
10	0.112899095460448\\
13.3333333333333	0.224876038494179\\
16.6666666666667	0.38874289687214\\
20	0.62109696162678\\
23.3333333333333	0.966270587480445\\
26.6666666666667	1.32807681821275\\
30	1.88750682644676\\
};\label{f5p1}
\addplot [color=black!50!green, dashed, line width=1.5pt]
  table[row sep=crcr]{%
0	0.00570188662260922\\
3.33333333333333	0.0136133737129793\\
6.66666666666667	0.0365459787102438\\
10	0.0745714351792469\\
13.3333333333333	0.142760688413568\\
16.6666666666667	0.261962264495615\\
20	0.463276211753134\\
23.3333333333333	0.733477864364371\\
26.6666666666667	1.03972774294224\\
30	1.45525095654133\\
};
\addlegendentry{Weighted MRT}

\addplot [color=black!50!green, dashed, line width=1.5pt, mark size=5.0pt, mark=o, mark options={solid, black!50!green}, forget plot]
  table[row sep=crcr]{%
0	0.00570188662260922\\
3.33333333333333	0.0136133737129793\\
6.66666666666667	0.0365459787102438\\
10	0.0745714351792469\\
13.3333333333333	0.142760688413568\\
16.6666666666667	0.261962264495615\\
20	0.463276211753134\\
23.3333333333333	0.733477864364371\\
26.6666666666667	1.03972774294224\\
30	1.45525095654133\\
};
\addplot [color=black!50!green, dashdotted, line width=1.5pt]
  table[row sep=crcr]{%
0	0.242528918404907\\
3.33333333333333	0.487188707154127\\
6.66666666666667	0.865962627006622\\
10	1.28017309000083\\
13.3333333333333	1.77262326156307\\
16.6666666666667	2.30440126583353\\
20	2.90844302896404\\
23.3333333333333	3.54598233365814\\
26.6666666666667	4.17037425641474\\
30	4.92292549277131\\
};
\addlegendentry{Optimal Digital}

\addplot [color=black!50!green, dashdotted, line width=1.5pt, mark size=5.0pt, mark=o, mark options={solid, black!50!green}, forget plot]
  table[row sep=crcr]{%
0	0.242528918404907\\
3.33333333333333	0.487188707154127\\
6.66666666666667	0.865962627006622\\
10	1.28017309000083\\
13.3333333333333	1.77262326156307\\
16.6666666666667	2.30440126583353\\
20	2.90844302896404\\
23.3333333333333	3.54598233365814\\
26.6666666666667	4.17037425641474\\
30	4.92292549277131\\
};
\addplot [color=black!50!green, dotted, line width=1.5pt]
  table[row sep=crcr]{%
0	0.198416833104121\\
3.33333333333333	0.403352259700429\\
6.66666666666667	0.744473760359661\\
10	1.11919694947664\\
13.3333333333333	1.6037640248776\\
16.6666666666667	2.11504298934294\\
20	2.69765819085419\\
23.3333333333333	3.30211728107045\\
26.6666666666667	3.88593701103228\\
30	4.53751605072873\\
};
\addlegendentry{Optimal Analog}

\addplot [color=black!50!green, dotted, line width=1.5pt, mark size=5.0pt, mark=o, mark options={solid, black!50!green}, forget plot]
  table[row sep=crcr]{%
0	0.198416833104121\\
3.33333333333333	0.403352259700429\\
6.66666666666667	0.744473760359661\\
10	1.11919694947664\\
13.3333333333333	1.6037640248776\\
16.6666666666667	2.11504298934294\\
20	2.69765819085419\\
23.3333333333333	3.30211728107045\\
26.6666666666667	3.88593701103228\\
30	4.53751605072873\\
};
\addplot [color=black, line width=1.5pt, mark size=3.5pt, mark=square, mark options={solid, black}, forget plot]
  table[row sep=crcr]{%
0	0.0178524587292708\\
3.33333333333333	0.0480762804458644\\
6.66666666666667	0.112560858157874\\
10	0.19741865413437\\
13.3333333333333	0.3602354816065\\
16.6666666666667	0.57975073558014\\
20	0.868470031784592\\
23.3333333333333	1.25814072032358\\
26.6666666666667	1.66427815861768\\
30	2.23296971000129\\
};
\addplot [color=black, line width=1.5pt, only marks, mark size=3.5pt, mark=square, mark options={solid, black}, forget plot]
  table[row sep=crcr]{%
0	0.0178524587292708\\
3.33333333333333	0.0480762804458644\\
6.66666666666667	0.112560858157874\\
10	0.19741865413437\\
13.3333333333333	0.3602354816065\\
16.6666666666667	0.57975073558014\\
20	0.868470031784592\\
23.3333333333333	1.25814072032358\\
26.6666666666667	1.66427815861768\\
30	2.23296971000129\\
};\label{f5p2}
\addplot [color=black, dashed, line width=1.5pt, mark size=3.5pt, mark=square, mark options={solid, black}, forget plot]
  table[row sep=crcr]{%
0	0.0113372405858569\\
3.33333333333333	0.0267512490276472\\
6.66666666666667	0.0687140988896127\\
10	0.134024393580386\\
13.3333333333333	0.242122910496644\\
16.6666666666667	0.413289442909186\\
20	0.674545158475804\\
23.3333333333333	0.997287047862083\\
26.6666666666667	1.36181862397583\\
30	1.8044044289565\\
};
\addplot [color=black, dashdotted, line width=1.5pt, mark size=3.5pt, mark=square, mark options={solid, black}, forget plot]
  table[row sep=crcr]{%
0	0.407979111319647\\
3.33333333333333	0.725047040429987\\
6.66666666666667	1.14632240343983\\
10	1.5939107859579\\
13.3333333333333	2.09990176905515\\
16.6666666666667	2.62698303625885\\
20	3.21418566071369\\
23.3333333333333	3.82817005829471\\
26.6666666666667	4.42732089883436\\
30	5.15339663249764\\
};
\addplot [color=black, dotted, line width=1.5pt, mark size=3.5pt, mark=square, mark options={solid, black}, forget plot]
  table[row sep=crcr]{%
0	0.341849439992547\\
3.33333333333333	0.618332710423623\\
6.66666666666667	1.00961607521543\\
10	1.42376510712176\\
13.3333333333333	1.9284337971619\\
16.6666666666667	2.44081680820209\\
20	3.01018741196915\\
23.3333333333333	3.59321754590557\\
26.6666666666667	4.15318448404502\\
30	4.77988802474256\\
};
\end{axis}

\node [draw,fill=white] at (rel axis cs: 0.26,0.8) {\shortstack[l]{
\ref{f5p1} $R_{0}^{\rm{lb}}$  \\
\ref{f5p2} $R_{0}$}};
\end{tikzpicture}%
}
    \caption{Tag rate versus transmit power.}\vspace{-5mm}
    \label{fig:rate_BD}
\end{figure}

We thus investigate the rates versus the BS transmit power under perfect SIC in Fig.~\ref{fig:rate_primary} and Fig.~\ref{fig:rate_BD}, respectively. {In Fig. \ref{fig:rate_BD}, $R_0^{\rm{lb}}$ is the lower bound rate from   \eqref{Eqn:upper}, and $R_0$ is the actual rate of the tag \eqref{Eqn:rate_BD}.} 

Fig.~\ref{fig:rate_primary} shows that without our designs, the symbiotic gains are negligible. For example,   with random beamforming, the user rates are the lowest. Also,  weighted MRT beamforming is computed without symbiotic goals.  It merely shapes the beam towards the users (ignoring the tag).  Then  $U_2$ achieves a low rate. However, $U_1$ achieves the highest rate because it performs SIC and cancels out the signal of $U_2$. On the other hand, with our designs, the two users achieve acceptable rates, exceeding the threshold rates, while also enhancing the harvested power and the rate performance of the tag (Fig.~\ref{fig:EH} and Fig.~\ref{fig:rate_BD}).  As observed, the rate performance of the tag significantly improves, and it achieves the highest rate while preserving the performance of the primary NOMA users -- Fig.~\ref{fig:rate_primary}.  We also see that digital beamforming achieves slightly better rate performance than analog, as it controls the phases and amplitudes by digital processing.  However, the former significantly decreases power consumption and hardware complexity while achieving acceptable performances.

If we compare  (a)  the SR network and (b) the equivalent NOMA system without the tag,  the two users achieve the highest rates for (b)   with proposed digital/analog beamforming designs. Network (b)  can also work with reduced transmit BS powers for certain rates of users.   For example, in (a), for  \qty{13.32}{bps/\Hz} and  \qty{7.6}{bps/\Hz} and \qty{2.1}{bps/\Hz} for $U_1$ and  $U_2$ and the tag (Fig.~\ref{fig:rate_BD}), respectively, BS requires \qty{13.33}{\dB m}. However, in (b), the required power decreases by $\sim$\qty{6}{\dB m} to achieve those rates for $U_1$  and $U_2$ ($U_1$ achieves  \qty{13.32}{bps/\Hz} at $p_t = \qty{6.27}{\dB m}$ and $U_2$ achieves  \qty{7.6}{bps/\Hz} at $p_t = \qty{8.33}{\dB m}$).  Thus,  symbiotic gains are not without costs. In (b), the user rates can be supported without considering the tag.  In (a), the BS must compensate for the deep fading of the tag channels and ensure that the tag will harvest enough.  

{Nonetheless, our algorithm ensures symbiosis has overall benefits as the rate requirements of both primary users and the tag can be satisfied and further enhanced.  Additionally, the tag harvests enough power for its activation.  Despite the fact that the SR+NOMA system has a slightly lower rate compared to the primary system without any tag in the low transmit power regime, e.g., $\sim \qty{96.3}{\percent}$ at $p_t = \qty{16}{\dB m}$, it surpasses the traditional NOMA network as the  transmit power increases, e.g., $\sim \qty{102.0}{\percent}$ at $p_t = \qty{30}{\dB m}$.}

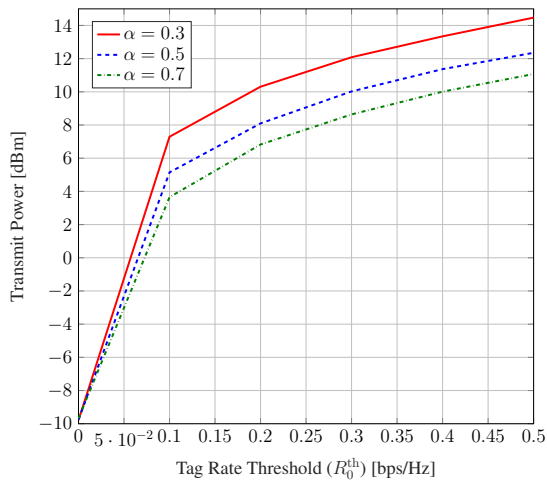
\begin{figure}[!t]\centering \vspace{-0mm}
    \fontsize{14}{14}\selectfont 
    \resizebox{.575\totalheight}{!}{
%
%
\begin{tikzpicture}

\begin{axis}[%
width=4.755in,
height=4.338in,
at={(0.798in,0.586in)},
scale only axis,
xmin=0,
xmax=0.5,
xlabel style={font=\color{white!15!black}},
xlabel={Tag Rate Threshold ($R^{\rm{th}}_0$) [bps/Hz]},
ymin=-10,
ymax=15,
ylabel style={font=\color{white!15!black}},
ylabel={Transmit Power [dBm]},
axis background/.style={fill=white},
xmajorgrids,
ymajorgrids,
legend style={at={(0.03,0.97)}, anchor=north west, legend cell align=left, align=left, draw=white!15!black}
]
\addplot [color=red, line width=1.5pt]
  table[row sep=crcr]{%
1e-07	-9.78145364310183\\
0.1	7.29277947417677\\
0.2	10.3090394837807\\
0.3	12.0840182778335\\
0.4	13.3427711087019\\
0.5	14.4711908594256\\
};
\addlegendentry{$\alpha=0.3$}

\addplot [color=blue, dashed, line width=1.5pt]
  table[row sep=crcr]{%
1e-07	-9.7623500475878\\
0.1	5.14614598592556\\
0.2	8.09774209318971\\
0.3	10.0244601544574\\
0.4	11.3674396722901\\
0.5	12.3511513235372\\
};
\addlegendentry{$\alpha=0.5$}

\addplot [color=black!50!green, dashdotted, line width=1.5pt]
  table[row sep=crcr]{%
1e-07	-9.68275339367909\\
0.1	3.65404429322064\\
0.2	6.82622429805531\\
0.3	8.63888564389902\\
0.4	10.0094162372516\\
0.5	11.0738572792338\\
};
\addlegendentry{$\alpha=0.7$}

\end{axis}
\end{tikzpicture}%
}
    \caption{The minimum transmit power as a function of $R_0^{\rm{th}}$.}\vspace{-1mm}
    \label{fig:minimum_required_power}
\end{figure}
\noindent \textbf{Transmit Power Minimization:}
Herein, we consider three  NOMA users ($K=3$) and investigate the TPMin problem (Section \ref{Sec:transmit_min}) with digital BS  transmit beamforming. Users  $U_k$, $k=1,2,3$, require the rates  $R_1^{\rm{th}}={\qty{2}{bps/\Hz}}$, $R_2^{\rm{th}}={\qty{1}{bps/\Hz}}$, and  $R_3^{\rm{th}}={\qty{0.5}{bps/\Hz}}$, respectively. We assume $d_f=\qty{3}{\m}$, $d_{h_k}=  \qty{10}{\m}$ ($k=1,2,3$), $d_{q_1}=\qty{8}{\m}$, $d_{q_2}=\qty{9}{\m}$, and $d_{q_3}=\qty{10}{\m}.$ These parameter values are taken from \cite{Xu2021NOMA, Zhuang2022}. 

\subsection{Does Symbiosis Work Better Under Low Transmit Powers?}
\begin{figure}[!t]\centering \vspace{-0mm}
    \fontsize{14}{14}\selectfont 
    \resizebox{.54\totalheight}{!}{\input{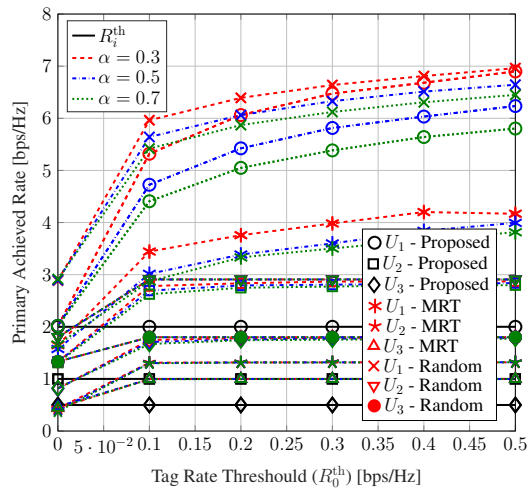}}
    \caption{User rates as a function of $R_0^{\rm{th}}$. }\vspace{-1mm}
    \label{fig:UserRate_Rth}
\end{figure}

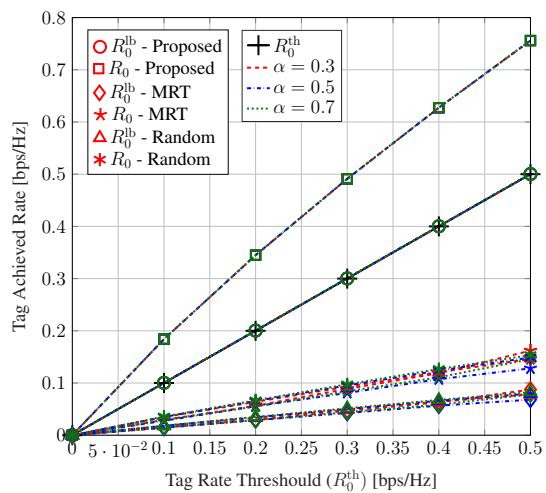
\begin{figure}[!t]\centering \vspace{-3mm}
    \fontsize{14}{14}\selectfont 
    \resizebox{.575\totalheight}{!}{
%
%
\begin{tikzpicture}

\begin{axis}[%
width=4.755in,
height=4.338in,
at={(0.798in,0.586in)},
scale only axis,
xmin=0,
xmax=0.5,
xlabel style={font=\color{white!15!black}},
xlabel={Tag Rate Threshould ($R^{\rm{th}}_0)$ [bps/Hz]},
ymin=0,
ymax=0.8,
ylabel style={font=\color{white!15!black}},
ylabel={Tag Achieved Rate [bps/Hz]},
axis background/.style={fill=white},
xmajorgrids,
ymajorgrids,
legend style={at={(0.37,0.97)}, anchor=north west, legend cell align=left, align=left, draw=white!15!black}
]
\addplot [color=black, line width=1.5pt, mark size=7.5pt, mark=+, mark options={solid, black}]
  table[row sep=crcr]{%
1e-07	9.99999999272475e-08\\
0.1	0.1\\
0.2	0.2\\
0.3	0.3\\
0.4	0.4\\
0.5	0.5\\
};
\addlegendentry{$R^{\rm{th}}_0$}

\addplot [color=red, dashed, line width=1.5pt]
  table[row sep=crcr]{%
1e-07	2.22654890255376e-05\\
0.1	0.100246423583745\\
0.2	0.200222300146645\\
0.3	0.300210664333172\\
0.4	0.400178579839323\\
0.5	0.500171999833757\\
};
\addlegendentry{$\alpha=0.3$}

\addplot [color=red, line width=1.5pt, only marks, mark size=4.5pt, mark=o, mark options={solid, red}, forget plot]
  table[row sep=crcr]{%
1e-07	2.22654890255376e-05\\
0.1	0.100246423583745\\
0.2	0.200222300146645\\
0.3	0.300210664333172\\
0.4	0.400178579839323\\
0.5	0.500171999833757\\
};\label{f8p1}
\addplot [color=blue, dashdotted, line width=1.5pt]
  table[row sep=crcr]{%
1e-07	4.5921650457842e-05\\
0.1	0.100293978203255\\
0.2	0.20042647449227\\
0.3	0.300381388836128\\
0.4	0.400283546981558\\
0.5	0.500326252199039\\
};
\addlegendentry{$\alpha=0.5$}

\addplot [color=blue, dashdotted, line width=1.5pt, mark size=4.5pt, mark=o, mark options={solid, blue}, forget plot]
  table[row sep=crcr]{%
1e-07	4.5921650457842e-05\\
0.1	0.100293978203255\\
0.2	0.20042647449227\\
0.3	0.300381388836128\\
0.4	0.400283546981558\\
0.5	0.500326252199039\\
};
\addplot [color=black!50!green, dotted, line width=1.5pt]
  table[row sep=crcr]{%
1e-07	5.33258514645112e-05\\
0.1	0.100314833783186\\
0.2	0.200412234189512\\
0.3	0.300491980811595\\
0.4	0.400444106534492\\
0.5	0.500428241358778\\
};
\addlegendentry{$\alpha=0.7$}

\addplot [color=black!50!green, dotted, line width=1.5pt, mark size=4.5pt, mark=o, mark options={solid, black!50!green}, forget plot]
  table[row sep=crcr]{%
1e-07	5.33258514645112e-05\\
0.1	0.100314833783186\\
0.2	0.200412234189512\\
0.3	0.300491980811595\\
0.4	0.400444106534492\\
0.5	0.500428241358778\\
};
\addplot [color=red, dashed, line width=1.5pt, mark size=3.5pt, mark=square, mark options={solid, red}, forget plot]
  table[row sep=crcr]{%
1e-07	0\\
0.1	0.183914112101279\\
0.2	0.344607287000846\\
0.3	0.490945924936071\\
0.4	0.627145327082462\\
0.5	0.755841401195178\\
};
\addplot [color=red, line width=1.5pt, only marks, mark size=3.5pt, mark=square, mark options={solid, red}, forget plot]
  table[row sep=crcr]{%
1e-07	0\\
0.1	0.183914112101279\\
0.2	0.344607287000846\\
0.3	0.490945924936071\\
0.4	0.627145327082462\\
0.5	0.755841401195178\\
};\label{f8p2}
\addplot [color=blue, dashdotted, line width=1.5pt, mark size=3.5pt, mark=square, mark options={solid, blue}, forget plot]
  table[row sep=crcr]{%
1e-07	0\\
0.1	0.183995006273483\\
0.2	0.344918873134605\\
0.3	0.491186237238548\\
0.4	0.627284020676516\\
0.5	0.756035063174139\\
};
\addplot [color=black!50!green, dotted, line width=1.5pt, mark size=3.5pt, mark=square, mark options={solid, black!50!green}, forget plot]
  table[row sep=crcr]{%
1e-07	0\\
0.1	0.184030481709404\\
0.2	0.344897143215961\\
0.3	0.49134189129535\\
0.4	0.627496152870347\\
0.5	0.756163101761013\\
};
\addplot [color=red, dashed, line width=1.5pt, forget plot]
  table[row sep=crcr]{%
1e-07	1.06901766268769e-05\\
0.1	0.0148957171601509\\
0.2	0.0293485534865604\\
0.3	0.0468353299766342\\
0.4	0.0631667295264484\\
0.5	0.0874445726269962\\
};
\addplot [color=red, line width=1.5pt, only marks, mark size=5.5pt, mark=diamond, mark options={solid, red}, forget plot]
  table[row sep=crcr]{%
1e-07	1.06901766268769e-05\\
0.1	0.0148957171601509\\
0.2	0.0293485534865604\\
0.3	0.0468353299766342\\
0.4	0.0631667295264484\\
0.5	0.0874445726269962\\
};\label{f8p3}
\addplot [color=blue, dashdotted, line width=1.5pt, mark size=5.5pt, mark=diamond, mark options={solid, blue}, forget plot]
  table[row sep=crcr]{%
1e-07	1.9390062177786e-05\\
0.1	0.0141882047346533\\
0.2	0.0288628085223419\\
0.3	0.042477983493305\\
0.4	0.0566148977749142\\
0.5	0.0680952649169929\\
};
\addplot [color=black!50!green, dotted, line width=1.5pt, mark size=5.5pt, mark=diamond, mark options={solid, black!50!green}, forget plot]
  table[row sep=crcr]{%
1e-07	2.47915397976281e-05\\
0.1	0.015082877762321\\
0.2	0.0286416811330517\\
0.3	0.0442541195552745\\
0.4	0.0586822520097357\\
0.5	0.0780869836509118\\
};
\addplot [color=red, dashed, line width=1.5pt, mark size=5.0pt, mark=star, mark options={solid, red}, forget plot]
  table[row sep=crcr]{%
1e-07	0\\
0.1	0.0290604694637206\\
0.2	0.0563541153845238\\
0.3	0.088744850224812\\
0.4	0.118847614244299\\
0.5	0.161953674201441\\
};
\addplot [color=red, line width=1.5pt, only marks, mark size=5.0pt, mark=star, mark options={solid, red}, forget plot]
  table[row sep=crcr]{%
1e-07	0\\
0.1	0.0290604694637206\\
0.2	0.0563541153845238\\
0.3	0.088744850224812\\
0.4	0.118847614244299\\
0.5	0.161953674201441\\
};\label{f8p4}
\addplot [color=blue, dashdotted, line width=1.5pt, mark size=5.0pt, mark=star, mark options={solid, blue}, forget plot]
  table[row sep=crcr]{%
1e-07	0\\
0.1	0.0277466135077833\\
0.2	0.0554222628871299\\
0.3	0.0810276238281886\\
0.4	0.107299426205864\\
0.5	0.128026804705294\\
};
\addplot [color=black!50!green, dotted, line width=1.5pt, mark size=5.0pt, mark=star, mark options={solid, black!50!green}, forget plot]
  table[row sep=crcr]{%
1e-07	0\\
0.1	0.0294618531352126\\
0.2	0.0550978638888931\\
0.3	0.0842300798606152\\
0.4	0.110921613299146\\
0.5	0.14566107985861\\
};
\addplot [color=red, dashed, line width=1.5pt, forget plot]
  table[row sep=crcr]{%
1e-07	1.24114125996621e-05\\
0.1	0.0178457412707757\\
0.2	0.0333832576531579\\
0.3	0.0490642807968677\\
0.4	0.0639321971948704\\
0.5	0.0784175271311486\\
};
\addplot [color=red, line width=1.5pt, only marks, mark size=5.0pt, mark=triangle, mark options={solid, red}, forget plot]
  table[row sep=crcr]{%
1e-07	1.24114125996621e-05\\
0.1	0.0178457412707757\\
0.2	0.0333832576531579\\
0.3	0.0490642807968677\\
0.4	0.0639321971948704\\
0.5	0.0784175271311486\\
};\label{f8p5}
\addplot [color=blue, dashdotted, line width=1.5pt, mark size=5.0pt, mark=triangle, mark options={solid, blue}, forget plot]
  table[row sep=crcr]{%
1e-07	2.07761592265865e-05\\
0.1	0.0179618287478026\\
0.2	0.0343646275467267\\
0.3	0.0503295577462377\\
0.4	0.0659276928823357\\
0.5	0.0798454271567814\\
};
\addplot [color=black!50!green, dotted, line width=1.5pt, mark size=5.0pt, mark=triangle, mark options={solid, black!50!green}, forget plot]
  table[row sep=crcr]{%
1e-07	2.96235457956403e-05\\
0.1	0.0183827267372729\\
0.2	0.0350307565465119\\
0.3	0.0515393396775162\\
0.4	0.0675260472405193\\
0.5	0.0825725392008171\\
};
\addplot [color=red, dashed, line width=1.5pt, mark size=5.0pt, mark=asterisk, mark options={solid, red}, forget plot]
  table[row sep=crcr]{%
1e-07	0\\
0.1	0.0346830449724617\\
0.2	0.0638154559854121\\
0.3	0.0930634939255288\\
0.4	0.120422778236753\\
0.5	0.146240189700584\\
};
\addplot [color=red, line width=1.5pt, only marks, mark size=5.0pt, mark=asterisk, mark options={solid, red}, forget plot]
  table[row sep=crcr]{%
1e-07	0\\
0.1	0.0346830449724617\\
0.2	0.0638154559854121\\
0.3	0.0930634939255288\\
0.4	0.120422778236753\\
0.5	0.146240189700584\\
};\label{f8p6}
\addplot [color=blue, dashdotted, line width=1.5pt, mark size=5.0pt, mark=asterisk, mark options={solid, blue}, forget plot]
  table[row sep=crcr]{%
1e-07	0\\
0.1	0.0349005251912018\\
0.2	0.06558645782867\\
0.3	0.0953342161212809\\
0.4	0.123989835010933\\
0.5	0.148738791463054\\
};
\addplot [color=black!50!green, dotted, line width=1.5pt, mark size=5.0pt, mark=asterisk, mark options={solid, black!50!green}, forget plot]
  table[row sep=crcr]{%
1e-07	0\\
0.1	0.0356799694466644\\
0.2	0.0667967449444741\\
0.3	0.097496159811389\\
0.4	0.126813630238984\\
0.5	0.153497074334574\\
};
\end{axis}

\begin{axis}[%
width=6.135in,
height=5.323in,
at={(0in,0in)},
scale only axis,
xmin=0,
xmax=1,
ymin=0,
ymax=1,
axis line style={draw=none},
ticks=none,
axis x line*=bottom,
axis y line*=left
]
\end{axis}

\node [draw,fill=white] at (rel axis cs: 0.275,0.76) {\shortstack[l]{
\ref{f8p1} $R_{0}^{\rm{lb}}$ - Proposed \\
\ref{f8p2} $R_{0}$ - Proposed \\
\ref{f8p3} $R_{0}^{\rm{lb}}$ - MRT \\
\ref{f8p4} $R_{0}$ - MRT \\
\ref{f8p5} $R_{0}^{\rm{lb}}$ - Random \\
\ref{f8p6} $R_{0}$ - Random}};

\end{tikzpicture}%
}
    \caption{Tag rate as a function of $R_0^{\rm{th}}$. }\vspace{-1mm}
    \label{fig:TagRate_Rth}
\end{figure}

The reduction of BS  power consumption is highly advantageous from an EE  standpoint as it allows for increased flexibility in connecting additional devices. However, it is essential to determine the minimum power requirement for the BS to support symbiosis. Additionally, understanding the impact of this power reduction on both user experience and tag performance is crucial. These critical issues are thoroughly investigated in Fig.~\ref{fig:minimum_required_power}, Fig.~\ref{fig:UserRate_Rth}, and Fig.~\ref{fig:TagRate_Rth}.

Fig.~\ref{fig:minimum_required_power} depicts the minimum transmit power versus the tag's rate threshold $R^{\rm{th}}_0$ for three $\alpha$ values under perfect SIC.  When the tag rate threshold reaches zero,  i.e., there is no tag, the network must maintain the minimum rates of the primary users. For example, at $p_t =-20\;$dBm,  the primary rate requirements are met (Fig.~\ref{fig:UserRate_Rth}). However, as the tag rate threshold increases, the BS needs more power to satisfy the increase.  For instance, for $\alpha = 0.5$, and the tag to reach  \qty{0.3}{bps/\Hz} while maintaining the primary rates and the tag EH constraint, the transmit power is \qty{10}{\dB m}. In contrast, it increases  by  \qty{2.4}{\dB m} for the tag to achieve  \qty{0.5}{bps/\Hz}. Moreover, for lower $\alpha$, the transmit power increases for a certain rate requirement of the tag. This is because when the tag reflects less power for communication, it can harvest more  (Section \ref{passive_tag}). 

Furthermore, we investigate the primary rates and the tag rate as functions of tag rate threshold $R^{\rm{th}}_0$ for different $\alpha$ in  Fig.~\ref{fig:UserRate_Rth} and Fig.~\ref{fig:TagRate_Rth}, respectively, under perfect SIC. We also investigate the rates of the devices when the BS adopts NRT and random beamforming. However,  these cannot fulfill the primary and tag rate requirements, so SR is not feasible. On the other hand, our beamforming design can not only maintain those requirements but also improve performance. Hence, our algorithm enables green SR under low transmit powers.

According to Fig.~\ref{fig:UserRate_Rth}, when the tag rate is reaching zero,  our design simply maintains the minimum rate requirements of the primary users. However, as $R^{\rm{th}}_0$ increases, our design ensures that the primary users achieve rates that exceed the respective threshold rates, while the tag achieves  $R_0^{\rm{lb}}$ (Fig.~\ref{fig:TagRate_Rth}). However,  the rate enhancement for $U_1$, whose effective channel is the best, is higher than those of $U_2$ and $U_3$, as $U_1$ is the user who performs SIC. For example,  for $R^{\rm{th}}_0 = {\qty{0.3}{bps/\Hz}}$ and $\alpha = 0.5$, $U_1$, $U_2$, and $U_3$ achieve the rates of \qty{5.9}{bps/\Hz}, \qty{2.9}{bps/\Hz} and \qty{1.85}{bps/\Hz} which are respectively \qty{3.9}{bps/\Hz}, \qty{1.9}{bps/\Hz}  and \qty{1.35}{bps/\Hz} higher than the respective threshold rates. The tag's actual rate $R_0$ is also higher than its threshold. In particular, for $R_0^{\rm{th}} = \qty{0.3}{bps/\Hz}$, the tag actually achieves  \qty{0.5}{bps/\Hz}.

We also observe that  TPMin ensures the rate requirements of the primary users regardless of the tag reflection levels,   $\alpha$.  However, for low reflection levels, $U_1$ achieves higher rates than for high $\alpha$ values. Specifically, $U_1$ achieves rate gains of \qty{31.3}{\percent} and \qty{15.6}{\percent} for $\alpha=\num{0.3}$ and $\alpha=\num{0.5}$, respectively, compared to $\alpha=\num{0.7}$ at $R^{\rm{th}}_0$ of \qty{0.3}{bps/\Hz}. This is because the BS increases the transmit power for low $\alpha$ values (Fig.~\ref{fig:minimum_required_power}).

Fig.~\ref{fig:minimum_required_power}, Fig.~\ref{fig:UserRate_Rth}, and Fig.~\ref{fig:TagRate_Rth} show that maintaining the tag rate requirement is important, i.e., the primary users exceed their rate thresholds, when the tag's rate is satisfied ($R_0^{\rm{lb}}$ in Fig.~\ref{fig:TagRate_Rth}).

\subsection{What is the Impact of SIC on Symbiosis? Do Our Algorithms Ease the Problem?}

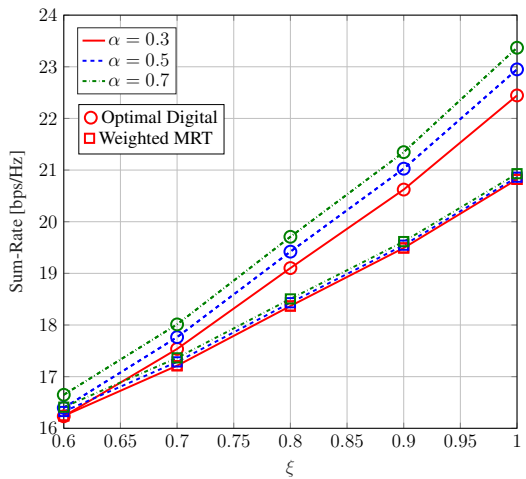
\begin{figure}[!t]\centering \vspace{-3mm}
    \fontsize{14}{14}\selectfont 
    \resizebox{.575\totalheight}{!}{
%
%
\begin{tikzpicture}

\begin{axis}[%
width=4.755in,
height=4.338in,
at={(0.798in,0.586in)},
scale only axis,
xmin=0.6,
xmax=1,
xlabel style={font=\color{white!15!black}},
xlabel={$\xi$},
ymin=16,
ymax=24,
ylabel style={font=\color{white!15!black}},
ylabel={Sum-Rate [bps/Hz]},
axis background/.style={fill=white},
xmajorgrids,
ymajorgrids,
legend style={at={(0.03,0.97)}, anchor=north west, legend cell align=left, align=left, draw=white!15!black}
]
\addplot [color=red, line width=1.5pt]
  table[row sep=crcr]{%
0.6	16.2326821376664\\
0.7	17.5363555644985\\
0.8	19.1021935892181\\
0.9	20.6224555763002\\
1	22.4450349243227\\
};
\addlegendentry{$\alpha=0.3$}

\addplot [color=red, line width=1.5pt, only marks, mark size=4.5pt, mark=o, mark options={solid, red}, forget plot]
  table[row sep=crcr]{%
0.6	16.2326821376664\\
0.7	17.5363555644985\\
0.8	19.1021935892181\\
0.9	20.6224555763002\\
1	22.4450349243227\\
};\label{f9p1}
\addplot [color=blue, dashed, line width=1.5pt]
  table[row sep=crcr]{%
0.6	16.3954068762646\\
0.7	17.7646998282224\\
0.8	19.4205336713137\\
0.9	21.0268655251839\\
1	22.9497298250707\\
};
\addlegendentry{$\alpha=0.5$}

\addplot [color=blue, dashed, line width=1.5pt, mark size=4.5pt, mark=o, mark options={solid, blue}, forget plot]
  table[row sep=crcr]{%
0.6	16.3954068762646\\
0.7	17.7646998282224\\
0.8	19.4205336713137\\
0.9	21.0268655251839\\
1	22.9497298250707\\
};
\addplot [color=black!50!green, dashdotted, line width=1.5pt]
  table[row sep=crcr]{%
0.6	16.6488111710479\\
0.7	18.0116230912691\\
0.8	19.7079650965648\\
0.9	21.3471118369276\\
1	23.3683998666639\\
};
\addlegendentry{$\alpha=0.7$}

\addplot [color=black!50!green, dashdotted, line width=1.5pt, mark size=4.5pt, mark=o, mark options={solid, black!50!green}, forget plot]
  table[row sep=crcr]{%
0.6	16.6488111710479\\
0.7	18.0116230912691\\
0.8	19.7079650965648\\
0.9	21.3471118369276\\
1	23.3683998666639\\
};
\addplot [color=red, line width=1.5pt, mark size=3.5pt, mark=square, mark options={solid, red}, forget plot]
  table[row sep=crcr]{%
0.6	16.2411795648575\\
0.7	17.2119061262083\\
0.8	18.3674938735802\\
0.9	19.4916884212039\\
1	20.8200198363877\\
};
\addplot [color=red, line width=1.5pt, only marks, mark size=3.5pt, mark=square, mark options={solid, red}, forget plot]
  table[row sep=crcr]{%
0.6	16.2411795648575\\
0.7	17.2119061262083\\
0.8	18.3674938735802\\
0.9	19.4916884212039\\
1	20.8200198363877\\
};\label{f9p2}
\addplot [color=blue, dashed, line width=1.5pt, mark size=3.5pt, mark=square, mark options={solid, blue}, forget plot]
  table[row sep=crcr]{%
0.6	16.3253727777866\\
0.7	17.2863487891318\\
0.8	18.4313870415561\\
0.9	19.5455863962656\\
1	20.8619545750803\\
};
\addplot [color=black!50!green, dashdotted, line width=1.5pt, mark size=3.5pt, mark=square, mark options={solid, black!50!green}, forget plot]
  table[row sep=crcr]{%
0.6	16.409251411524\\
0.7	17.3641283171295\\
0.8	18.505242137532\\
0.9	19.6155153443038\\
1	20.9299926423691\\
};
\end{axis}

\begin{axis}[%
width=6.135in,
height=5.323in,
at={(0in,0in)},
scale only axis,
xmin=0,
xmax=1,
ymin=0,
ymax=1,
axis line style={draw=none},
ticks=none,
axis x line*=bottom,
axis y line*=left
]
\end{axis}

\node [draw,fill=white] at (rel axis cs: 0.28,0.7) {\shortstack[l]{
\ref{f9p1} Optimal Digital \\
\ref{f9p2} Weighted MRT}};

\end{tikzpicture}%
}
    \caption{The  sum-rate as a function  of $\xi$ for $p_t=\qty{20}{\dB m}$.}\vspace{-1mm}
    \label{fig:sumRate_imperfect}
\end{figure}
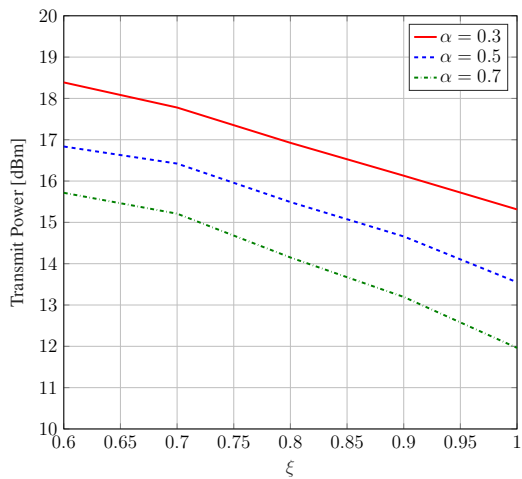
\begin{figure}[!t]\centering \vspace{-0mm}
    \fontsize{14}{14}\selectfont 
    \resizebox{.575\totalheight}{!}{
%
%
\begin{tikzpicture}

\begin{axis}[%
width=4.755in,
height=4.338in,
at={(0.798in,0.586in)},
scale only axis,
xmin=0.6,
xmax=1,
xlabel style={font=\color{white!15!black}},
xlabel={$\xi$},
ymin=10,
ymax=20,
ylabel style={font=\color{white!15!black}},
ylabel={Transmit Power [dBm]},
axis background/.style={fill=white},
xmajorgrids,
ymajorgrids,
legend style={legend cell align=left, align=left, draw=white!15!black}
]
\addplot [color=red, line width=1.5pt]
  table[row sep=crcr]{%
0.6	18.3871982145279\\
0.7	17.7795207730633\\
0.8	16.9251333165361\\
0.9	16.1297779679626\\
1	15.3138683325526\\
};
\addlegendentry{$\alpha=0.3$}

\addplot [color=blue, dashed, line width=1.5pt]
  table[row sep=crcr]{%
0.6	16.8358737672323\\
0.7	16.422768081425\\
0.8	15.4916496170187\\
0.9	14.6594213167472\\
1	13.550694072475\\
};
\addlegendentry{$\alpha=0.5$}

\addplot [color=black!50!green, dashdotted, line width=1.5pt]
  table[row sep=crcr]{%
0.6	15.7155820956831\\
0.7	15.2103927125604\\
0.8	14.1502976989677\\
0.9	13.1952568272584\\
1	11.9621689086921\\
};
\addlegendentry{$\alpha=0.7$}

\end{axis}

\begin{axis}[%
width=6.135in,
height=5.323in,
at={(0in,0in)},
scale only axis,
xmin=0,
xmax=1,
ymin=0,
ymax=1,
axis line style={draw=none},
ticks=none,
axis x line*=bottom,
axis y line*=left
]
\end{axis}
\end{tikzpicture}%
}
    \caption{The minimum transmit power as a function of $\xi$.}\vspace{-1mm}
    \label{fig:PtxDelta}
\end{figure}

To answer these questions, we consider the performance of  WSRMax and TPMin with  SIC decoding errors. To this end, we plot the sum rate and the minimum transmit power as a function of the imperfect SIC coefficient ($\xi$) for various tag reflection coefficients, $\alpha$, in Fig. \ref{fig:sumRate_imperfect} and Fig. \ref{fig:PtxDelta}. There are   two primary  users ($K=2$) and for the TPMin algorithm, we set  $R_1^{\rm{th}}={\qty{3}{bps/\Hz}}$,  $R_2^{\rm{th}}={\qty{1}{bps/\Hz}}$, and $R_0^{\rm{th}}={\qty{0.5}{bps/\Hz}}$ for $U_1$, $U_2$, and the tag, respectively.

According to Fig.~\ref{fig:sumRate_imperfect}, the sum rate increases with the SIC quality,  $\xi\rightarrow 1$, i.e., perfect SIC. Conversely,  SIC imperfection degrades symbiosis. For instance, with our beamforming and power allocation, at $p_t=\qty{20}{\dB m}$ and $\alpha= 0.5$, we observe a  {\qty{1.92}{bps/\Hz}} loss for $\xi = 0.9$, compared to the perfect SIC case.  Nonetheless, our algorithms, i.e., Algorithm \ref{Algo1} and Algorithm \ref{Algo2}, outperform weighted MRT beamforming, and the resulting rate enhancement is more noticeable as $\xi$ approaches \num{1}.

Fig. \ref{fig:PtxDelta} also shows that, as  $\xi\rightarrow 0$, i.e., severe SIC imperfection,  the BS needs more power to overcome the detrimental effects of SIC imperfection and maintain the minimum rate of $U_k$ and tag. However, as  $\xi$ approaches $1$, i.e., perfect SIC,  less power is required to maintain the rates.  For instance, for $\alpha = 0.5$ and $\xi = 1$, \qty{13.5}{\dB m} transmit power is required to meet the device rate requirements, while it increases by \qty{2.8}{\dB m} for $\xi = 0.7$.

Therefore, the success of symbiosis is significantly influenced by imperfections in SIC, highlighting the crucial need for successful SIC decoding. However, our proposed algorithms effectively mitigate the impact of imperfect SIC, ensuring that the performance trends remain relatively stable regardless of the value of $\xi$. These results provide valuable insights for improving the performance of imperfect SIC/SR  in various scenarios.

\subsection{Convergence Rate of the Proposed Algorithms}\label{sec:convergence}

\begin{figure}[!t]\centering\vspace{-3mm}
    \fontsize{14}{14}\selectfont 
    \resizebox{.575\totalheight}{!}{
%
%
\begin{tikzpicture}

\begin{axis}[%
width=4.755in,
height=4.338in,
at={(0.798in,0.586in)},
scale only axis,
xmin=1,
xmax=10,
xlabel style={font=\color{white!15!black}},
xlabel={Iteration Number},
ymin=4.5,
ymax=7.5,
ylabel style={font=\color{white!15!black}},
ylabel={Objective Value},
axis background/.style={fill=white},
xmajorgrids,
ymajorgrids,
legend style={at={(0.97,0.03)}, anchor=south east, legend cell align=left, align=left, draw=white!15!black}
]
\addplot [color=blue, line width=1.5pt]
  table[row sep=crcr]{%
1	4.87059847572755\\
2	5.36409209615105\\
3	5.49633380680294\\
4	5.59633380680294\\
5	5.59633380680294\\
6	5.59633380680294\\
7	5.59633380680294\\
8	5.59633380680294\\
9	5.59633380680294\\
10	5.59633380680294\\
11	5.59633380680294\\
12	5.59633380680294\\
13	5.59633380680294\\
14	5.59633380680294\\
15	5.59633380680294\\
16	5.59633380680294\\
17	5.59633380680294\\
18	5.59633380680294\\
19	5.59633380680294\\
20	5.59633380680294\\
};
\addlegendentry{$p_t=10$ dBm}

\addplot [color=red, dashed, line width=1.5pt]
  table[row sep=crcr]{%
1	5.91134621716061\\
2	6.10232303799048\\
3	6.20347296182777\\
4	6.20347296182777\\
5	6.20347296182777\\
6	6.20347296182777\\
7	6.20347296182777\\
8	6.20347296182777\\
9	6.20347296182777\\
10	6.20347296182777\\
11	6.20347296182777\\
12	6.20347296182777\\
13	6.20347296182777\\
14	6.20347296182777\\
15	6.20347296182777\\
16	6.20347296182777\\
17	6.20347296182777\\
18	6.20347296182777\\
19	6.20347296182777\\
20	6.20347296182777\\
};
\addlegendentry{$p_t=15$ dBm}

\addplot [color=black!50!green, dashdotted, line width=1.5pt]
  table[row sep=crcr]{%
1	6.61295424814991\\
2	6.91295424814991\\
3	7.02991230124989\\
4	7.02991230124989\\
5	7.02991230124989\\
6	7.02991230124989\\
7	7.02991230124989\\
8	7.02991230124989\\
9	7.02991230124989\\
10	7.02991230124989\\
11	7.02991230124989\\
12	7.02991230124989\\
13	7.02991230124989\\
14	7.02991230124989\\
15	7.02991230124989\\
16	7.02991230124989\\
17	7.02991230124989\\
18	7.02991230124989\\
19	7.02991230124989\\
20	7.02991230124989\\
};
\addlegendentry{$p_t=20$ dBm}

\end{axis}

\begin{axis}[%
width=6.135in,
height=5.323in,
at={(0in,0in)},
scale only axis,
xmin=0,
xmax=1,
ymin=0,
ymax=1,
axis line style={draw=none},
ticks=none,
axis x line*=bottom,
axis y line*=left
]
\end{axis}
\end{tikzpicture}%
}
    \caption{The convergence of the WSRMax objective value.}
    \label{fig:objctVal}\vspace{-1mm}
\end{figure}
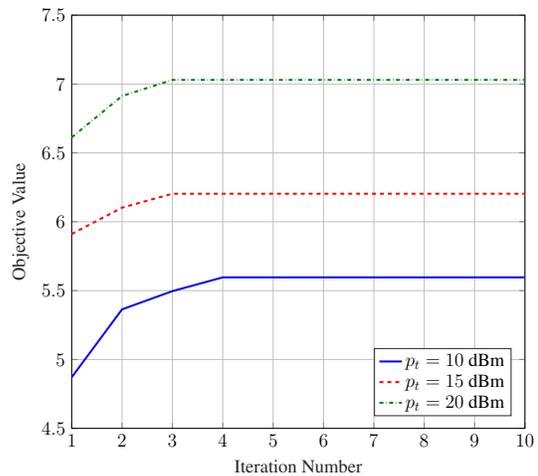

\begin{figure}[!t]\centering
    \fontsize{14}{14}\selectfont 
    \resizebox{.575\totalheight}{!}{
%
%
\begin{tikzpicture}

\begin{axis}[%
width=4.755in,
height=4.338in,
at={(0.798in,0.586in)},
scale only axis,
xmin=1,
xmax=10,
xlabel style={font=\color{white!15!black}},
xlabel={Iteration Number},
ymin=3,
ymax=9,
ylabel style={font=\color{white!15!black}},
ylabel={Objective Value},
axis background/.style={fill=white},
xmajorgrids,
ymajorgrids,
legend style={at={(0.97,0.03)}, anchor=south east, legend cell align=left, align=left, draw=white!15!black}
]
\addplot [color=blue, line width=1.5pt]
  table[row sep=crcr]{%
1	3.1243807143327\\
2	4.65429543695859\\
3	4.65429543695859\\
4	4.65429543695859\\
5	4.65429543695859\\
6	4.65429543695859\\
7	4.65429543695859\\
8	4.65429543695859\\
9	4.65429543695859\\
10	4.65429543695859\\
11	4.65429543695859\\
12	4.65429543695859\\
13	4.65429543695859\\
14	4.65429543695859\\
15	4.65429543695859\\
16	4.65429543695859\\
17	4.65429543695859\\
18	4.65429543695859\\
19	4.65429543695859\\
20	4.65429543695859\\
};
\addlegendentry{$R^{\rm{th}}_0=0.2$ bps/Hz}

\addplot [color=red, dashed, line width=1.5pt]
  table[row sep=crcr]{%
1	4.17384889487559\\
2	7.00911181590249\\
3	7.00911181590249\\
4	7.00911181590249\\
5	7.00911181590249\\
6	7.00911181590249\\
7	7.00911181590249\\
8	7.00911181590249\\
9	7.00911181590249\\
10	7.00911181590249\\
11	7.00911181590249\\
12	7.00911181590249\\
13	7.00911181590249\\
14	7.00911181590249\\
15	7.00911181590249\\
16	7.00911181590249\\
17	7.00911181590249\\
18	7.00911181590249\\
19	7.00911181590249\\
20	7.00911181590249\\
};
\addlegendentry{$R^{\rm{th}}_0=0.3$ bps/Hz}

\addplot [color=black!50!green, dashdotted, line width=1.5pt]
  table[row sep=crcr]{%
1	5.47506814706099\\
2	8.64711764431723\\
3	8.64711764431723\\
4	8.64711764431723\\
5	8.64711764431723\\
6	8.64711764431723\\
7	8.64711764431723\\
8	8.64711764431723\\
9	8.64711764431723\\
10	8.64711764431723\\
11	8.64711764431723\\
12	8.64711764431723\\
13	8.64711764431723\\
14	8.64711764431723\\
15	8.64711764431723\\
16	8.64711764431723\\
17	8.64711764431723\\
18	8.64711764431723\\
19	8.64711764431723\\
20	8.64711764431723\\
};
\addlegendentry{$R^{\rm{th}}_0=0.4$ bps/Hz}

\end{axis}

\begin{axis}[%
width=6.135in,
height=5.323in,
at={(0in,0in)},
scale only axis,
xmin=0,
xmax=1,
ymin=0,
ymax=1,
axis line style={draw=none},
ticks=none,
axis x line*=bottom,
axis y line*=left
]
\end{axis}
\end{tikzpicture}%
}
    \caption{The convergence of the TPMin objective value.}
    \label{fig:ObjectivePM}\vspace{-1mm}
\end{figure}
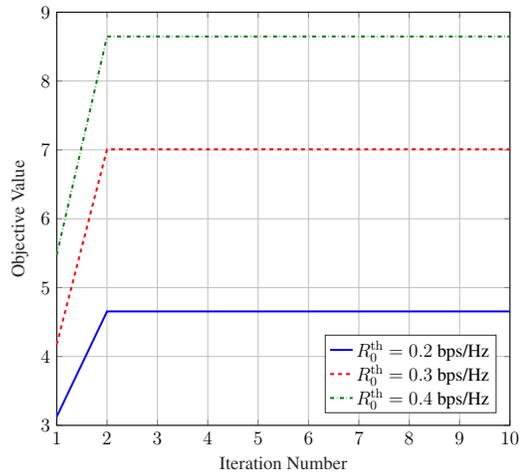 

Fig. \ref{fig:objctVal} and Fig. \ref{fig:ObjectivePM} investigate the convergence rates of the proposed  WSRMax algorithm (Section \ref{sec:WSM}, Remark \ref{rem:overal_algo_WSR})  and TPMin algorithm (Section \ref{Sec:transmit_min}, Remark \ref{rem:overal_algo_PM}), respectively.

Fig. \ref{fig:objctVal}   shows the convergence behavior of the former for two different BS  transmit power levels, i.e., $p_t=\{10,15,20\}\,\text{dBm}$. The objective function of the overall algorithm is the WSR. The stopping condition for convergence is that the increment of the normalized objective function is less than $\epsilon=10^{-3}$. As shown in Fig. \ref{fig:objctVal}, the WSR, obtained by combining Algorithms 1 and 2,  increases rapidly and saturates as the number of outer loop iterations increases. Specifically, the overall algorithm converges in less than 4 iterations regardless of the BS transmit power.  

Fig. \ref{fig:ObjectivePM} also shows the convergence behavior of \textbf{Algorithm 3} for three different rate thresholds. We use the same stopping criteria. It converges very quickly.

\section{Conclusion}\label{conclusion}

Symbiotic radio networks are rapidly emerging as a promising solution for providing extensive connectivity to passive IoT devices. In this study, we have focused on such a network between a primary NOMA  network and an \abc tag, where we aim to optimize the system performance and power efficiency. To achieve this, we have designed an optimal beamformer and power allocation scheme for the BS  that maximizes the WSR  or minimizes the transmit power of the BS while satisfying the minimum rate requirements of both the users and the tag. Our formulation takes into account the EH  constraint of the tag and the imperfect SIC decoding.  

Since these optimization problems are inherently non-convex, we have developed solution algorithms based on alternating optimization, fixed-point, and semi-definite relaxation techniques. By leveraging these algorithms, we have achieved significant performance improvements in the BS designs while ensuring that the rate requirements are met. Notably, our approach does not require any modifications to the tag, thereby preserving the cost benefits associated with passive tags.

The proposed symbiotic radio network framework enables efficient collaboration between the primary NOMA network and the \abc tag, facilitating massive connectivity for passive IoT devices. Through the optimized BS designs and power allocation, we have achieved enhanced system performance and power efficiency while considering practical constraints such as tag EH and imperfect SIC. Here are some of the insights gained.
\begin{enumerate}

    \item Our algorithms deliver more power and rates for the secondary network  (Fig.~\ref{fig:EH}). For example,  for $p_t=20$\qty{}{\dB m,}  $M=32$, and  digital beamforming,  the tag achieves  \qty{4.52}{\dB m} and  \qty{3.98}{bps/\Hz}, respectively. Without our algorithms, i.e., with random beamforming,  the tag cannot exceed \qty{0.2}{\dB m} and \qty{0.96}{bps/\Hz}. Thus, our designs significantly outperform random beamforming.

    \item Our algorithms can enable symbiosis while satisfying the rate constraints and the EH requirements. For example,  for $p_t=20$\qty{}{\dB m,}  $M=32$, and  digital beamforming at the BS,  $U_1$, $U_2$, and the tag achieve  \qty{15}{bps/\Hz}, \qty{8.1}{bps/\Hz}, and \qty{3.2}{bps/\Hz}, respectively (Fig.~\ref{fig:rate_primary} and Fig.~\ref{fig:rate_BD}), with $\alpha = {0.6}$ at the tag. Moreover, the  tag can harvest  \qty{3}{\dB m} (Fig.~\ref{fig:EH}).

    \item Symbiosis is also enabled with analog beamforming, which significantly reduces the BS  power consumption and the hardware complexity at a small loss of the rates.  For example,   with a  \num{32}-antenna BS at  \qty{20}{\dB m},   $U_1$, $U_2$, and the tag loss \qty{0.74}{bps/\Hz}, \qty{0.3}{bps/\Hz}, and \qty{0.21}{bps/\Hz} (Fig.~\ref{fig:rate_primary} and Fig.~\ref{fig:rate_BD}). However, the BS needs just one RF chain for analog beamforming but \num{32}  for digital beamforming. 
    
    \item Our algorithms achieve symbiosis while reducing power consumption and keeping the performance of both networks.  For example,  with $\alpha = 0.5$ at the tag, in order that $U_1$, $U_2$, $U_3$, and the tag achieve \qty{5.81} {bps/\Hz}, \qty{2.81}{bps/\Hz}, \qty{1.75}{bps/\Hz}, and \qty{0.49}{bps/\Hz}, respectively, with the  BS just at \qty{10}{\dB m} (Fig.~\ref{fig:minimum_required_power}). 
    
   \item  SIC imperfection degrades the efficacy of symbiosis. For instance,  for $\alpha = 0.5$ and $\xi = 0.9$, it degrades the sum rate and power by {\qty{1.92}{bps/\Hz}} (Fig.~\ref{fig:sumRate_imperfect}) and \qty{1.11}{\dB m} (Fig.~\ref{fig:PtxDelta}). Nonetheless,  WSRMax and TPMin ameliorate such losses. 
    
\end{enumerate}

Although we derived new algorithms,   there are several possible improvements. First, we assumed the availability of  CSI, which provides a performance upper bound.    CSI  estimation is challenging with multiple tags and users. Thus, the impact of imperfect CSI on symbiosis should be investigated. Second, our algorithm could be extended for the nonlinear EH models.  Third, symbiosis may involve more than just one tag.  However,  multiple tags create additional interference terms, and optimal algorithms are thus needed.   
    \vspace{-4mm}
 \appendices
 \section{Feasibility problem for $\mathbf{w}$ \eqref{P1w1_prob}}\label{apx:Convexify}
According to \eqref{P1w1_prob}, in order to solve the feasibility sub-problem for finding $\mathbf{w}$, the SINR constraints of the primary users and tag \eqref{P1w1_rateUi}, are respectively convexified as  $\frac{1}{\sqrt{\gamma_{k}^{\rm{th}}}} \sqrt{\rho_k p_t} \vert \mathbf{h}_i^{\rm{H}} \mathbf{w} \vert \geq \Vert \mathbf{v}_i^k \Vert$ and $ \frac{1}{2\sqrt{\gamma_{0}^{\rm{th}}}} \sqrt{\alpha p_t}  \vert q_1 \vert \vert \mathbf{f}_1^{\rm{H}} \mathbf{w} \vert \geq \Vert \mathbf{v}_0 \Vert$,
where  $\mathbf{v}_i^k = [\mathbf{v}_{i,k}^1,  \mathbf{v}_{i,k}^2, \sqrt{\alpha p_t }  \vert \mathbf{g}_i^{\rm{H}} \mathbf{w} \vert,  \sigma]$, in which $\mathbf{v}_{i,k}^1 = \sqrt{p_t} \vert \mathbf{h}_i^{\rm{H}} \mathbf{w} \vert [\sqrt{\rho_1}, \ldots,  \sqrt{\rho_{k-1}}]$ and $\mathbf{v}_{i,k}^2 = \sqrt{p_t} \vert \mathbf{h}_i^{\rm{H}} \mathbf{w} \vert[\sqrt{\left( 2-2 \xi_{k+1} \right) \rho_{k+1}}, \ldots,  \sqrt{\left( 2-2 \xi_{K} \right) \rho_{K}}]$. Besides, $\mathbf{v}_0 = [\mathbf{v}_{0}^1,  \sigma]$, in which ~$\mathbf{v}_{0}^1 = \sqrt{p_t} \vert \mathbf{h}_1^{\rm{H}} \mathbf{w}  \vert[ \sqrt{\rho_1  \left( 2-2 \xi_1 \right)}, \ldots, \sqrt{\rho_K \left( 2-2 \xi_K \right)}]$. Moreover, \eqref{P1w1_BD_EH} is realized by using the first-order Taylor
approximation as $p_b/\eta_b(1-\alpha)p_t \leq\mathbf{w}^{(i-1)} \mathbf{F} \mathbf{w}^{(i-1)} +\left( (\mathbf{F}+\mathbf{F}^{\rm{H}}) \mathbf{w}^{(i-1)}\right)^{\rm{H}} \left( \mathbf{w}^{(i)} -\mathbf{w}^{(i-1)}\right)$, 
where  $\mathbf{F} \triangleq \mathbf{f} \mathbf{f}^{\rm{H}}$. Besides, $\mathbf{w}^{(i)}$ and $\mathbf{w}^{(i-1)}$ are the current and the previous iteration values of $\mathbf{w}$.

\linespread{1.0}

\bibliographystyle{IEEEtran}
\bibliography{IEEEabrv,Ref}

\end{document}